\documentclass{article}

\usepackage[utf8]{inputenc}
\usepackage[T1]{fontenc}

\PassOptionsToPackage{table, dvipsnames}{xcolor}
\usepackage{xcolor}

\PassOptionsToPackage{colorlinks=true, citecolor=blue, linkcolor=blue, urlcolor=black}{hyperref}

\usepackage{arxiv}
\usepackage[utf8]{inputenc}
\usepackage[T1]{fontenc}
\usepackage{url}
\usepackage{booktabs}
\usepackage{nicefrac}
\usepackage{microtype}
\usepackage{lipsum}
\usepackage{graphicx}
\usepackage{natbib}
\usepackage{doi}
\usepackage{algorithm}
\usepackage{algpseudocode}%
\usepackage{listings}%
\usepackage{multirow}
\usepackage{multicol}
\usepackage{amsmath,amssymb,amsfonts}
\usepackage{amsthm}
\usepackage{mathrsfs}
\usepackage{float}
\usepackage{bm}
\usepackage{caption}
\usepackage{hyperref}

\usepackage{tabularx}
\usepackage{array}
\usepackage{pdflscape}

\title{Depth-progressive 3D seismic velocity model building via 2D generative diffusion models}

\author{
	\href{https://orcid.org/0000-0001-8868-7967}{\includegraphics[scale=0.06]{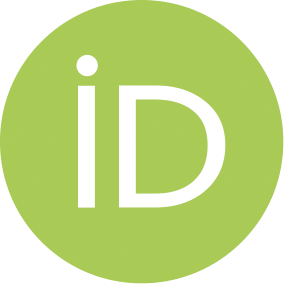}\hspace{1mm}Shijun~Cheng} \\
	Division of Physical Science and Engineering\\
	King Abdullah University of Science and Technology\\
	Thuwal 23955-6900, Saudi Arabia \\
        \And
    {Randy Harsuko} \\
	Division of Physical Science and Engineering\\
	King Abdullah University of Science and Technology\\
	Thuwal 23955-6900, Saudi Arabia \\
        \And
    {Mohammad H. Taufik} \\
	Division of Physical Science and Engineering\\
	King Abdullah University of Science and Technology\\
	Thuwal 23955-6900, Saudi Arabia \\
        \And
    {Tariq Alkhalifah} \\
	Division of Physical Science and Engineering\\
	King Abdullah University of Science and Technology\\
	Thuwal 23955-6900, Saudi Arabia \\
    [3ex]
  $^{*}$Corresponding author: \textbf{Shijun Cheng}~(\texttt{sjcheng.academic@gmail.com})
}

\renewcommand{\shorttitle}{Depth-progressive 3D VMB}

\hypersetup{
pdftitle={A template for the arxiv style},
pdfsubject={q-bio.NC, q-bio.QM},
pdfauthor={David S.~Hippocampus, Elias D.~Striatum},
pdfkeywords={First keyword, Second keyword, More},
}

\begin{document}
\maketitle

\begin{abstract}
Three-dimensional (3D) velocity model building (VMB) remains computationally challenging with conventional approaches and native 3D neural networks, largely due to the prohibitive memory and training cost of full 3D formulations. Therefore, we introduce a depth-progressive framework that achieves 3D VMB using 2D generative diffusion models through a depth-as-channel formulation, in which the depth dimension of each 3D velocity patch is mapped to the channel dimension of a 2D U-Net. This allows the model to synthesize overlapping depth patches sequentially from shallow to deep while conditioning the prediction of the deeper patch on the shallower one, as well as on any available geophysical constraints, including well logs and structural attributes. During training, paired shallow-deep patch pairs with randomized overlaps are used to teach the network how deeper velocity structures relate to shallower information, and a classifier-free guidance strategy enables the model to handle missing constraints gracefully. At inference, starting from a shallow velocity prior, the model recursively generates deeper patches in a top-down manner, and a depth-attenuated Gaussian blending scheme merges overlapping predictions into spatially coherent 3D volumes while accounting for the progressive accumulation of uncertainty at greater depths. Numerical experiments on both in-distribution benchmark models, including Overthrust, SEAM Arid, and SEG/EAGE, and the out-of-distribution Marmousi model demonstrate that the method effectively propagates shallow velocity priors downward and produces high-quality 3D velocity volumes across a range of geological settings and conditioning scenarios. Compared to a 3D benchmark of equivalent architecture, the proposed method significantly reduces parameter count, training cost, and inference cost, while maintaining superior VMB accuracy. These results demonstrate a practical and scalable alternative to full 3D architectures for large-scale VMB.
\end{abstract}

\keywords{Velocity model building \and 3D \and  Generative diffusion model \and Depth-progressive}
\section{\textbf{Introduction}}
Accurate subsurface velocity models are a prerequisite for reliable seismic imaging \citep{etgen2009overview}, and the subsurface is three-dimensional (3D). Lateral velocity variations caused by geological complexity can be severe, and a two-dimensional (2D) model built along a single profile will inevitably misposition reflectors and introduce migration artifacts when the geology deviates from the assumed symmetry \citep{gray2001seismic}. Three-dimensional (3D) velocity model building (VMB) addresses this problem by estimating a volumetric velocity field that accounts for out-of-plane variations, and the quality of downstream products such as depth-migrated images and amplitude-versus-offset analysis depend directly on how well this 3D model is resolved \citep{jones2010introduction}. As exploration has moved into increasingly complex geological settings, particularly in deep-water environments with rugose salt and steep structural dips, the need for accurate 3D velocity estimation has become more urgent \citep{petmecky2009improving}. Modern 3D seismic surveys now cover vast areas with dense spatial sampling, which adds an additional challenge: the volume of data to be processed, and the spatial extent of the model to be estimated, have grown to a point where computational burden is itself a limiting factor in the choice of a velocity analysis strategy \citep{ben2008velocity, virieux2009overview}.

Classical approaches to 3D VMB are largely adaptations of methods originally developed for two dimensions. Reflection tomography and migration velocity analysis update a smooth background velocity by minimizing moveout residuals measured in common-image-point gathers \citep{scales1987tomographic, symes2008migration}. These methods have become standard practice in industry and can handle large 3D datasets, but they depend on picking coherent reflection events across a range of offsets and azimuths, which becomes difficult in areas of poor illumination or strong lateral heterogeneity. Slope tomography \citep{lambare2008stereotomography} reduces the reliance on manual picking by working with locally measured dip information, though it shares the fundamental limitation of recovering only long-wavelength velocity structure. 3D full-waveform inversion (FWI) makes the most complete use of the recorded data by iteratively minimizing the misfit between observed and modeled waveforms \citep{virieux2009overview, warner2013anisotropic, metivier2016optimal}. Given a sufficiently accurate starting model, FWI can recover velocity variations approaching a resolution on the order of the dominant wavelength. The cost, however, is considerable: each FWI iteration requires numerical simulation of the seismic wavefield over the entire 3D model volume, and for exploration-scale problems this translates to millions of core hours per run. Beyond the computational burden, FWI is sensitive to the quality of the initial model. When the starting velocity is too far from the true solution, waveform residuals exceed half a period and cycle skipping drives the inversion toward spurious local minima \citep{bunks1995multiscale}. For these reasons, 3D FWI remains a specialized tool rather than a routine workflow component in many exploration settings.

Data-driven methods have attracted considerable interest as an alternative to these physics-based workflows. Convolutional neural networks (CNNs) trained on synthetic data have been shown to predict 2D velocity models directly from seismic shot gathers or stacked sections, replacing iterative inversion with a single forward pass through a trained network \citep{araya2018deep, wu2019inversionnet, yang2019deep, li2019deep, du2022deep}. Subsequent studies have incorporated physical constraints, attention mechanisms, and recurrent connections to improve the generalization of these methods to more complex geology \citep{sun2020theory, sun2021physics}. Generative models have also been applied to the velocity estimation problem: variational autoencoders and generative adversarial networks have been used to introduce learned geological priors that regularize the solution \citep{laloy2018training, mosser2020stochastic}. Among recent generative approaches, diffusion models have shown particular promise due to their training stability and their ability to produce diverse, high-quality samples from complex distributions \citep{ho2020denoising, song2020score}. For example, \cite{wang2023prior} introduced a diffusion-prior-regularized FWI framework in which a pretrained diffusion model was used to constrain acoustic velocity updates, and this idea was further extended by \cite{taufik2024learned} to multi-parameter elastic FWI, where diffusion-based learned regularization helps preserve structural consistency among elastic parameters and mitigate parameter cross-talk. Unlike diffusion-prior-regularized FWI methods that incorporate learned priors directly into the inversion updates, \cite{zhang2025well} used a diffusion model to build reliable initial velocity models constrained by well logs and structural information, thereby improving the starting point for subsequent FWI. Extending any of these approaches to 3D, however, introduces a steep increase in computational demand. A 3D CNN processes volumetric data with cubic growth in both memory and parameter count relative to its 2D counterpart, requiring large training datasets and GPU resources that are often impractical to assemble. This difficulty has limited most published neural network methods for VMB to two-dimensional problems, and a scalable approach for 3D velocity estimation at realistic exploration scales remains an open problem \citep{zeng2021inversionnet3d, rusmanugroho20223d}.

We recently proposed a depth-progressive strategy for 2D VMB using conditional diffusion models \citep{chengdiffvmbpart1, chengdiffvmbpart2}. For Earth surface recorded data, we decomposed the velocity model construction into a sequence of local prediction steps ordered from shallow to deep, rather than attempting to recover the entire model in a single operation. A conditional diffusion model is trained to synthesize a deep velocity patch given the overlying shallow patch as a condition, with optional inputs from sparse well-log velocities and structural attributes derived from seismic images. During training, these auxiliary constraints are randomly withheld following a classifier-free guidance scheme \citep{ho2022classifier}, so the model remains functional when constraints are absent at inference time. Prediction is performed recursively: starting from a known or estimated shallow velocity, the model generates successive depth patches, each of which becomes the conditioning input for the next step. This recursive mechanism from shallow to deep reflects the inherent nature of propagation of information in our seismic inverse problem for surface recorded data, where near-surface velocity is generally better constrained than deeper structure. Applied to 2D profiles, the method produced geologically plausible models with good agreement with well and structural constraints across a range of synthetic and field test cases.

In this work, we extend this depth-progressive approach to 3D VMB while keeping the computational footprint of a 2D diffusion model. Rather than training memory-intensive 3D networks, we apply 2D diffusion models in a depth-progressive manner, constructing velocity volumes from shallow to deep by explicitly propagating shallow velocity information into deeper regions. Specifically, we treat the depth extent of each 3D velocity patch as the channel dimension of the 2D CNN, a design we refer to as the depth-as-channel formulation. Each lateral slice of the 3D patch occupies one channel, and the network learns the relationships among slices at different depths through standard 2D convolutions over the channel axis, thereby avoiding 3D convolutions and reducing memory consumption and training time. The recursive inference scheme carries over from the 2D framework without modification, and can be conditioned on available geophysical constraints, including sparse well-log velocities and structural attributes, enabling geologically plausible VMB. By decomposing the 3D problem into a series of 2D synthesis tasks along the depth direction and merging predictions through patch fusion, our method achieves computational speedups comparable to 3D workflows. We evaluate the method on a suite of 3D synthetic velocity models and compare it against a full 3D diffusion approach, demonstrating that the depth-as-channel formulation recovers high-quality 3D velocity structures at substantially lower computational burden.
\section{\textbf{Prior work: Depth-progressive 2D velocity model building}}\label{sec:review}

Diffusion models learn a data distribution by reversing a gradual noising process. In the forward process, clean data $x_0$ drawn from the target distribution $q(x_0)$ is progressively corrupted by adding Gaussian noise while attenuating the signal over $T$ discrete timesteps:
\begin{equation}\label{eq1}
    q(x_t \mid x_{t-1}) = \mathcal{N}\!\left(x_t;\,
    \sqrt{1 - \beta_t}\, x_{t-1},\, \beta_t \mathbf{I}\right),
\end{equation}
where $\{\beta_t\}_{t=1}^T$ is a fixed variance schedule. By defining $\bar{\alpha}_t = \prod_{s=1}^t (1 - \beta_s)$, the noised sample at an arbitrary timestep can be written in closed form as:
\begin{equation}\label{eq2}
    x_t = \sqrt{\bar{\alpha}_t}\,x_0
    + \sqrt{1 - \bar{\alpha}_t}\,\epsilon, \quad
    \epsilon \sim \mathcal{N}(\mathbf{0}, \mathbf{I}).
\end{equation}
As $t \to T$, the distribution of $x_t$ approaches a standard Gaussian with zero mean. The reverse process learns to recover clean data from the noisy samples. In practice, the denoising network can be parameterized to predict either the added noise $\epsilon$ or the clean data $x_0$ directly. Both parameterizations are equivalent in principle, as the two quantities are related through equation~(2). In our implementation, we adopt the $x_0$-prediction parameterization, training a 2D U-Net $f_\theta(x_t, t)$ with the objective:
\begin{equation}\label{eq3}
    \mathcal{L} = \mathbb{E}_{x_0,\, \epsilon,\, t}
    \left\| x_0 - f_\theta(x_t, t) \right\|^2.
\end{equation}
New samples are generated by starting from $x_T \sim \mathcal{N}(\mathbf{0}, \mathbf{I})$ and iteratively applying the denoising network until a clean sample is recovered.

The central idea of our depth-progressive VMB framework is to decompose the construction of a full velocity model into a sequence of local prediction tasks ordered from shallow to deep. Rather than recovering the entire velocity volume in a single inversion, the model learns the conditional relationship between a shallow velocity patch and the deeper patch immediately below it. This decomposition is geologically motivated: due to the nature of sedimentation, as well as faults and folds, geological features are expected to have some form of connection along the depth direction, and since our data are recorded on the surface, conditioning the prediction of the deep on shallow velocity information naturally encodes this continuity. Building on this idea, we train a conditional diffusion model whose target distribution is the deep velocity patch $v_\text{deep}$, conditioned on several sources of information. The primary spatial condition is the overlying shallow velocity patch $v_\text{shallow}$. To make the model aware of the absolute depth position of each patch within the model, we additionally provide the depth position grids of both patches, denoted by $d_\text{shallow} \in \mathbb{R}^{n_z \times n_x}$ and $d_\text{deep} \in \mathbb{R}^{n_z \times n_x}$, where $n_z$ and $n_x$ are the spatial dimensions of a patch. These four inputs are concatenated along the channel dimension to form the network input $\{x_t, v_\text{shallow}, d_\text{shallow}, d_\text{deep}\}$, where $x_t$ denotes the noised version of $v_\text{deep}$ at diffusion timestep $t$. Beyond these spatial conditions, the model also accepts two optional geophysical constraints: sparse well-log velocities $w$ and structural attributes $s$ derived from seismic images such as reflectivity or a migrated image. The conditional training objective is:
\begin{equation}\label{eq4}
    \mathcal{L} = \mathbb{E}_{v_\text{deep},\, \boldsymbol{\epsilon},\, t}
    \left\| v_\text{deep} - f_\theta\!\left(x_t,\, t,\,
    v_\text{shallow},\, d_\text{shallow},\, d_\text{deep},\, w,\, s\right) \right\|^2.
\end{equation}
 
In practice, well logs and structural attributes are not always available at inference time, and the model must remain functional when some or all of these constraints are absent. To address this, we follow a classifier-free guidance strategy \citep{ho2022classifier} and mix conditional and unconditional training within a single run. At each training iteration, $w$ and $s$ are independently set to $\varnothing$ (None) with a fixed dropout probability, indicating to the network that the corresponding constraint is unavailable. This encourages the model to learn both the fully conditioned distribution $p(v_\text{deep} \mid v_\text{shallow}, d_\text{shallow}, d_\text{deep}, w, s)$ and its partially conditioned variants without requiring separate training runs. The depth position grids and the shallow velocity patch are always provided, as they form the essential spatial context for the depth-progressive construction. Training pairs $(v_\text{shallow}, v_\text{deep})$ are extracted from a set of velocity models by sliding a depth window of size $n_z$ along the depth axis. Rather than using a fixed stride, the overlap between each shallow and deep patch pair is drawn randomly from the range $[1, n_z-1]$ at each extraction, so that the model is exposed to a wide variety of relative depth offsets between the two patches during training. The corresponding depth position grids $d_\text{shallow}$ and $d_\text{deep}$ are computed from the absolute depth coordinates of each extracted patch pair.
 
Once trained, the model constructs a full velocity model through a recursive inference procedure. Let the model domain span depths $z_0$ to $z_{\text{max}}$, and let each patch cover a depth interval of size $\Delta z$. Consecutive patches overlap by $\delta z$ in the depth direction, so that the effective depth stride between successive predictions is $\Delta z - \delta z$. The inference begins with a shallow velocity prior $\hat{v}_\text{deep}^{(0)}$ (equivalent to $v_\text{shallow}$ in equation \ref{eq4}) at the topmost depth interval, obtained from conventional velocity analysis or available well information. At step $k$, the model generates the next patch by running the reverse diffusion chain conditioned on the previously predicted patch:
\begin{equation}\label{eq5}
    \hat{v}_\text{deep}^{(k)} = f_\theta\!\left(x_T,\, \hat{v}_\text{deep}^{(k-1)},\,
    d_\text{deep}^{(k-1)},\, d_\text{deep}^{(k)},\, w^{(k)},\, s^{(k)}\right),
    \quad x_T \sim \mathcal{N}(\mathbf{0}, \mathbf{I}),
\end{equation}
where $d_\text{deep}^{(k)}$ is the depth position grid for the $k$-th step, and $w^{(k)}$ and $s^{(k)}$ denote the well and structural constraints projected onto the corresponding depth interval. The predicted patch $\hat{v}_\text{deep}^{(k)}$ is then used as the shallow condition $v_\text{shallow}$ for step $k+1$, and the procedure continues until the depth $z_{\text{max}}$ is reached. Because consecutive patches overlap in the depth direction, the overlapping region receives predictions from both the current and the previous step. To merge these predictions without introducing discontinuities or boundary artifacts, we apply Gaussian-weighted blending \citep{shi2024generative} in the overlap region. Specifically, each predicted patch is weighted by a Gaussian envelope $g(z)$ that peaks at the patch center and decays toward its boundaries, and the final velocity at each depth sample is obtained as the normalized weighted average of all patches contributing to that location:
\begin{equation}\label{eq6}
    \hat{v}(z) = \frac{\sum_k g^{(k)}(z)\, \hat{v}_\text{deep}^{(k)}(z)}
    {\sum_k g^{(k)}(z)},
\end{equation}
where the sum runs over all patches whose depth range includes $z$. This blending scheme ensures a smooth transition between successive predictions along the depth axis. For models with large lateral extent, the same Gaussian blending is applied in the horizontal directions as well, where overlapping lateral patches are processed independently and merged in the same manner.

\section{\textbf{3D velocity model building with 2D diffusion models}}\label{sec:method}
Building on the depth-progressive framework described in Section~\ref{sec:review}, we extend it to 3D VMB while retaining the efficiency of a 2D diffusion model. The key idea is to treat the depth dimension of each 3D velocity patch as the channel dimension of the 2D U-Net, avoiding 3D convolutions entirely. This is preferred as the objective is to predict/forecast features along the depth direction, and this is better handled by the channel representation of such a direction, and confining the feature extraction to the depth slice \citep{borovykh2017conditional}. In the following subsections, we describe the input representation, the conditioning scheme, the recursive 3D inference procedure, and the network architecture.

\subsection{Depth-as-Channel representation and network input}\label{sec:method:sub1}

Extending the depth-progressive framework to 3D in a straightforward way would require replacing the 2D diffusion model with a 3D counterpart, where the denoising network operates directly on volumetric patches. While conceptually simple, this approach is prohibitively expensive, as 3D convolutions multiply the memory footprint and parameter count substantially compared to their 2D equivalents, making it difficult to train and deploy such a network at practical patch sizes. To avoid this cost while retaining the ability to model 3D velocity structure, we adopt a depth-as-channel formulation in which the depth dimension of each 3D velocity patch is mapped to the channel dimension of a 2D convolutional network.

Specifically, each 3D velocity patch has dimensions $n_z \times n_y \times n_x$, where $n_z$ is the number of depth samples and $n_x$, $n_y$ are the lateral dimensions. Under the depth-as-channel mapping, the patch is treated as a 2D feature map of size $n_x \times n_y$ with $n_z$ channels, so that standard 2D convolutions operate over the lateral plane while the depth structure is encoded across channels. Since the number of channels in a convolutional network is fixed by the architecture, $n_z$ must be set at training time, where we use $n_z = 16$ in our implementation. The lateral dimensions $n_x$ and $n_y$ are in principle unconstrained, but larger lateral extents increase memory consumption, so we set $n_x = n_y = 128$ in our following experiments.

With this representation in place, the network input is formed by concatenating two components along the channel dimension. The first is the noised deep velocity patch $x_t \in \mathbb{R}^{n_z \times n_y \times n_x}$, obtained by corrupting the target $v_\text{deep}$ according to the forward diffusion process at timestep $t$ (i.e., using Equation~(\ref{eq2})), and reshaped under the depth-as-channel convention. The second is the shallow velocity patch $v_\text{shallow} \in \mathbb{R}^{n_z \times n_y \times n_x}$, which is first passed through a lightweight input convolution layer (described in subsection \ref{sec:method:sub3}) before being concatenated with $x_t$. The absolute depth position of the patch pair is not supplied as an explicit input channel. Instead, the absolute depth position of the patch pair is represented by a single scalar $d_\text{max}$, the depth coordinate of the deepest grid point in $v_\text{deep}$. This simplification is made possible by fixing the overlap between consecutive shallow and deep patches at $n_z / 2$ depth samples: with a constant relative offset between the two patches, $d_\text{max}$ alone is sufficient to fully determine the absolute depth position of the entire patch pair. Rather than supplying $d_\text{max}$ as an explicit input channel, it is embedded into the network in a manner analogous to the diffusion timestep $t$ and injected into the residual blocks of the U-Net at each resolution stage. The details of this embedding are described in Section~\ref{sec:method:sub3}.

In addition to the depth position, the network can also incorporate available geophysical constraints to guide the prediction toward geologically plausible solutions. Structural attributes $s \in \mathbb{R}^{n_z \times n_y \times n_x}$, such as reflectivity or a migrated seismic section cropped to the patch extent, are handled in the same way as the velocity patches: the depth dimension is mapped to the channel dimension, producing a 2D tensor of size $n_y \times n_x$ with $n_z$ channels. This tensor is injected into the residual blocks of the U-Net at each resolution stage, allowing structural information to influence the prediction at multiple spatial scales. When structural attributes are unavailable, $s$ is set to $\varnothing$ (None) following the classifier-free dropout strategy described in Section~\ref{sec:review}.

Well-log velocities provide point location measurements of the true velocity along the depth axis. In practice, wells are expensive to drill, so we consider the limited but realistic scenario of a single well within the model domain. To incorporate this information into the network, we expand the 1D well velocity profile $w$ into a dense array of the same spatial dimensions as $v_\text{deep}$ by replicating the well values along the lateral directions. The expanded array is then injected into the residual blocks of the U-Net at each resolution stage alongside the structural attributes. To inform the network of the lateral position of the well within the patch, the well location coordinates are embedded in a manner analogous to the diffusion timestep $t$ and injected into the residual blocks at each stage. The details of this embedding are described in Section~\ref{sec:method:sub3}. When no well data are available, $w$ is set to $\varnothing$ following the same classifier-free dropout strategy.

\subsection{3D recursive inference with depth-attenuated Gaussian blending}
\label{sec:method:sub2}
At inference time, the 3D velocity model is constructed by the same recursive procedure used in the 2D framework, extended to handle the additional lateral dimension. The model domain is first partitioned into overlapping 3D patches of size $n_z \times n_y \times n_x$ along all three spatial axes. Along the depth axis, consecutive patches overlap by $n_z / 2$ samples, consistent with the fixed overlap used during training. Along the lateral axes, the overlap is chosen to ensure that the entire model volume is covered without gaps.

Inference proceeds from shallow to deep. The process begins with a shallow velocity prior $\hat{v}_\text{deep}^{(0)}$ covering the topmost $n_z$ depth samples, which may be obtained from conventional velocity analysis or available well information. At step $k$, the model takes $\hat{v}_\text{deep}^{(k-1)}$ as the shallow condition and generates the next patch by running the reverse diffusion chain:
\begin{equation}\label{eq7}
    \hat{v}_\text{deep}^{(k)} = f_\theta\!\left(x_T,\, \hat{v}_\text{deep}^{(k-1)},\,
    d_\text{max}^{(k)},\, w^{(k)},\, s^{(k)}\right),
    \quad x_T \sim \mathcal{N}(\mathbf{0}, \mathbf{I}),
\end{equation}
where $d_\text{max}^{(k)}$ is the depth coordinate of the deepest grid point in the $k$-th patch, and $w^{(k)}$ and $s^{(k)}$ are the well and structural constraints projected onto the corresponding depth interval. The predicted patch $\hat{v}_\text{deep}^{(k)}$ is passed forward as the shallow condition for step $k+1$, and the recursion continues until the full depth extent of the model is covered.

Once all depth patches along a given lateral position have been predicted, they are assembled along the depth axis through a depth-attenuated Gaussian blending scheme that accounts for the progressively increasing uncertainty in deeper predictions. Each predicted patch $\hat{v}_\text{deep}^{(k)}$ is weighted by a Gaussian envelope centered at the patch midpoint $z_k^c$ with standard deviation $\sigma_z = n_z / 8$:
\begin{equation}\label{eq8}
    g_z^{(k)}(z) = a^{(k)} \cdot \exp\!\left(-\frac{(z - z_k^c)^2}{2\sigma_z^2}\right),
\end{equation}
where $a^{(k)}$ is a depth-dependent attenuation factor that decreases linearly from one at the shallowest patch to a minimum value $w_{min}$ at the deepest patch:
\begin{equation}\label{eq9}
    a^{(k)} = 1 - (1 - w_{min}) \cdot \frac{k}{N - 1},
\end{equation}
with $N$ denoting the total number of depth patches, and $w_{min}$ is set to 0.1 in the following experiments. This attenuation reflects the physical reality that errors introduced at shallow depths propagate downward through the recursive inference, causing predictions in deeper regions to carry greater uncertainty. The velocity at each depth sample is then obtained as the normalized weighted sum over
all patches covering that depth:
\begin{equation}\label{eq10}
    \hat{v}(z) = \frac{\sum_k g_z^{(k)}(z)\, \hat{v}_\text{deep}^{(k)}(z)}
    {\sum_k g_z^{(k)}(z)}.
\end{equation}

Since diffusion models do not reliably generalize to lateral dimensions significantly larger than those seen during training, surveys whose lateral extent exceeds the patch size $n_x \times n_y$ are handled by partitioning the horizontal plane into overlapping patches and carrying out the depth-progressive inference independently for each lateral
patch. The resulting depth-assembled volumes are then merged in the lateral directions using a 2D Gaussian weight centered at the horizontal midpoint of each patch with standard deviation $\sigma_{x} = n_x / 8$ and $\sigma_{y} = n_y / 8$:
\begin{equation}\label{eq11}
    g_{xy}(x, y) = \exp\!\left(-\frac{(x - x_c)^2 + (y - y_c)^2}{2\sigma_{xy}^2}
    \right).
\end{equation}
The final 3D velocity volume is obtained by accumulating the depth-fused prediction from all lateral patches, weighted by $g_{xy}$, and normalizing by the total lateral weight. This two-stage blending suppresses boundary artifacts along all three spatial axes while naturally down-weighting less reliable predictions in deeper regions.

\begin{figure*}[htbp]
\centering
\includegraphics[width=0.9\textwidth]{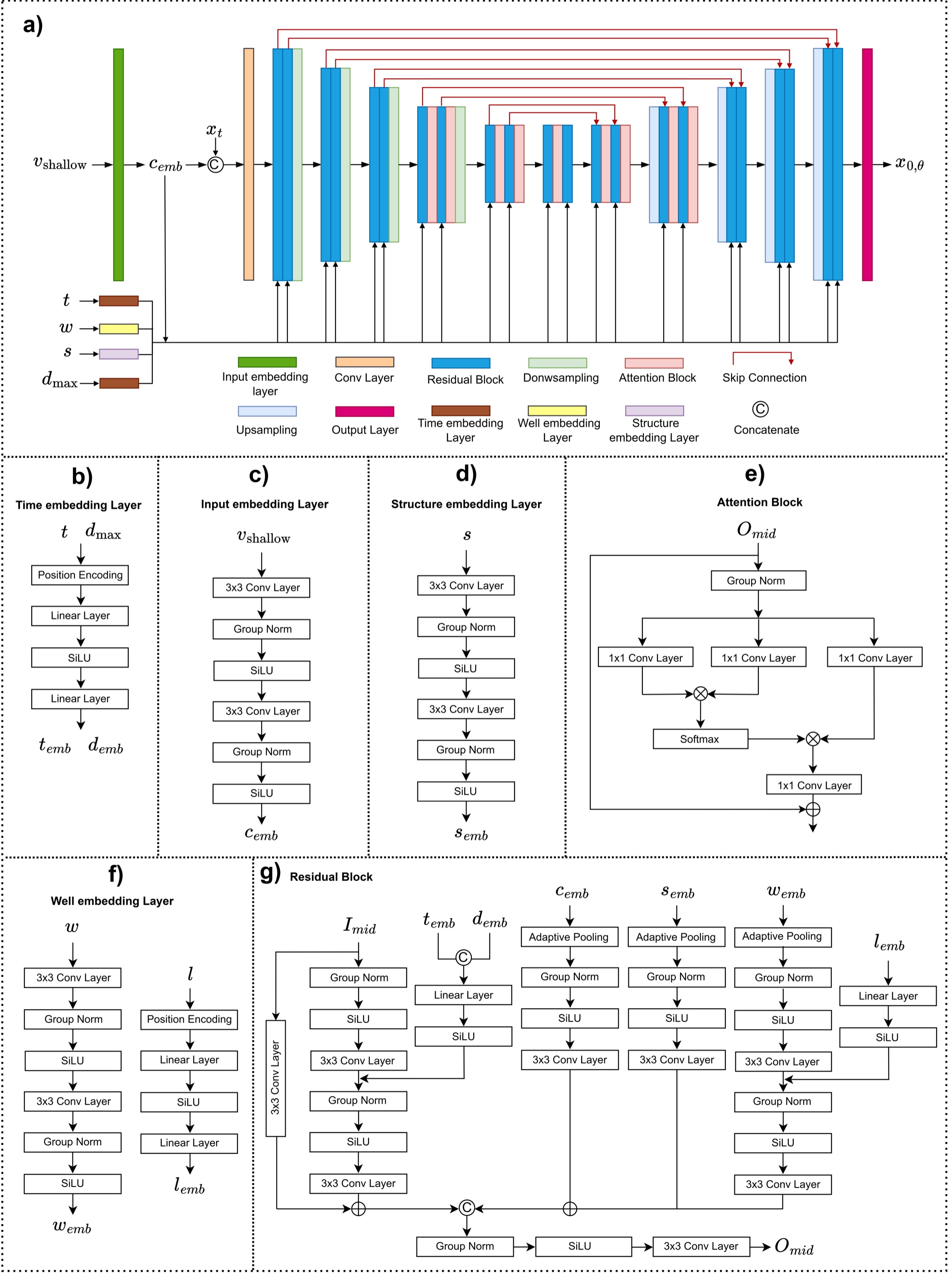}
\caption{An illustration of the proposed network architecture. (a) The overall network structure. (b) Time embedding layer (where $t$ is the diffusion timestep and $d_{max}$ is the depth coordinate of the deepest grid point in $v_{deep}$). (c) Input embedding layer (where $v_\text{shallow}$ is the shallow velocity patch). (d) Structure embedding layer (where $s$ is the structural attribute). (e) The attention block (where $O_{mid}$ is the block output in (g)). (f) Well embedding layer (where $w$ is the well velocity and $l$ is the lateral position of the well). (g) The residual block (where $I_{mid}$ is the block input and $O_{mid}$ is the block output).}
\label{fig1}
\end{figure*}

\subsection{Network architecture}\label{sec:method:sub3}

The denoising network follows a U-Net architecture with four encoding and four decoding stages, as illustrated in Figure~\ref{fig1}a. The feature map channel dimensions at the five stages are 64, 128, 256, 512, and 1024, respectively. Downsampling between stages is performed by stride-2 $3\times3$ convolutions, while upsampling uses nearest neighbor interpolation followed by a $3\times3$ convolution. Skip connections link each encoding stage to its corresponding decoding stage. Attention blocks are applied at the deeper stages of the network to capture long-range spatial dependencies.

The shallow velocity patch $v_\text{shallow} \in \mathbb{R}^{n_z \times n_y \times n_x}$, represented under the depth-as-channel convention, is first processed by an input embedding layer consisting of two blocks of $3\times3$ convolution, group normalization (GN), and sigmoid linear unit (SiLU) activation, which projects it from $n_z = 16$ channels to 64 channels and produces the embedding $c_{emb}$. This embedding serves two roles: it is concatenated with the noised input $x_t$ to form the network input, and it is also passed into the residual blocks at each stage to provide continuous shallow velocity context throughout the network.

The diffusion timestep $t$ and the patch depth coordinate $d_\text{max}$ are each embedded independently using the same structure: a sinusoidal position encoding followed by two linear layers with a SiLU activation in between, producing embeddings $t_{emb}$ and $d_{emb}$, respectively. These two embeddings are concatenated and passed through a linear layer and SiLU activation before being injected into the residual blocks at every stage.

The structural attribute $s \in \mathbb{R}^{n_z \times n_y \times n_x}$, reshaped under the depth-as-channel convention, is processed by a structure embedding layer composed of two blocks of $3\times3$ convolution, GN, and SiLU activation, yielding $s_{emb}$. The well velocity profile $w$, expanded into a dense array of the same dimensions as $v_\text{deep}$, is processed by an analogous well embedding layer to produce $w_{emb}$. The lateral position of the well within the patch, given as a 2D coordinate $l = (x, y)$, is embedded using sinusoidal position encoding followed by two linear layers with a SiLU activation, producing $l_{emb}$.

Each residual block takes an intermediate feature map $I_{mid}$ and incorporates all conditioning signals, as shown in Figure~\ref{fig1}g. The main processing branch applies GN, SiLU activation, and two sequential $3\times3$ convolution layers. The concatenated time and depth embedding $(t_{emb} \copyright\ d_{emb})$ is projected by a linear layer with SiLU activation and added to the intermediate feature map after the first convolution. The shallow velocity embedding $c_{emb}$, structural embedding $s_{emb}$, and well embedding $w_{emb}$ are each spatially adapted to the current feature map resolution through adaptive pooling, followed by GN, SiLU activation, and a $3\times3$ convolution; their contributions are summed and added to the main branch. The well position embedding $l_{emb}$ is projected through a linear layer with SiLU activation and likewise added to the feature map. The result of the main branch and the combined conditioning branch are concatenated, then passed through GN, SiLU activation, and a $3\times3$ convolution that halves the channel dimension to produce the output $O_{mid}$.

The attention block, applied at the deeper stages of the U-Net, operates on the intermediate feature map $O_{mid}$. After GN, three parallel $1\times1$ convolutions produce query, key, and value representations. The attention weights are computed by a softmax over the dot product of query and key, and the weighted value is aggregated and passed through a final $1\times1$ convolution. The result is added to the input via a residual connection.
\section{\textbf{Numerical examples}}\label{sec:examples}
We test the proposed method on four 3D velocity models with different levels of structural complexity. The Overthrust, SEAM Arid, and SEG/EAGE examples are used as in-distribution samples, because their test volumes are generated from unseen slices of the same source models used to construct part of the training set. The Marmousi model is used as an out-of-distribution sample to examine the generalization ability of the trained model to a different geological style. However, in using generative models to help regularize the inversion problem, we always hope that the true solution is within the training sample distribution as such distribution act as prior. This is accomplished by embedding our experience and prior knowledge of the subsurface as much as possible in the training set. For each in and out of distribution case, we evaluate the predicted 3D velocity volume both visually and quantitatively, and compare the proposed depth-as-channel 2D diffusion model with a 3D convolutional benchmark. We first describe the training data, augmentation strategy, conditioning information, and training configuration. We then present the results for each test model, followed by a quantitative comparison using mean absolute error (MAE) and structural similarity index (SSIM).

\begin{figure}[!htb] 
\centering
\includegraphics[width=1\textwidth]{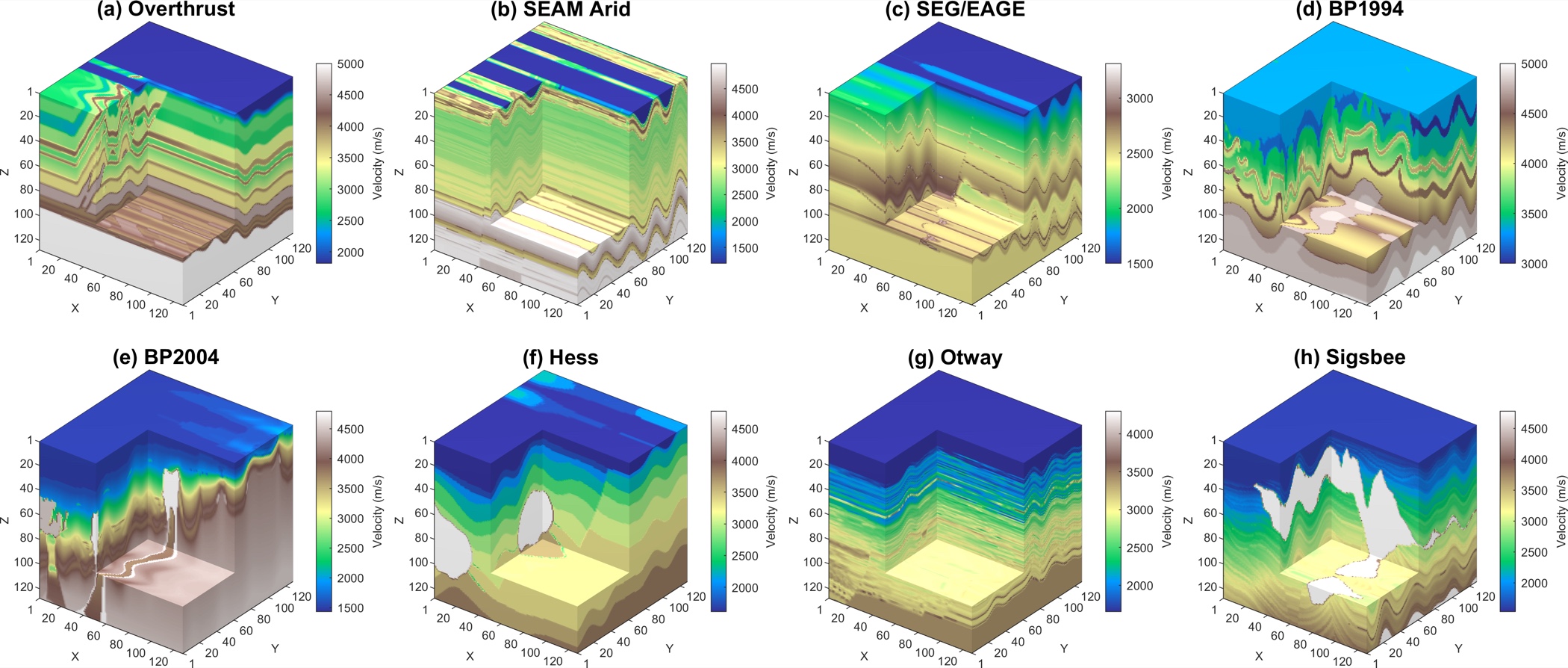}
\caption{Representative training samples from each source model.}
\label{fig2}
\end{figure}

\begin{figure}[!htb] 
\centering
\includegraphics[width=0.66\textwidth]{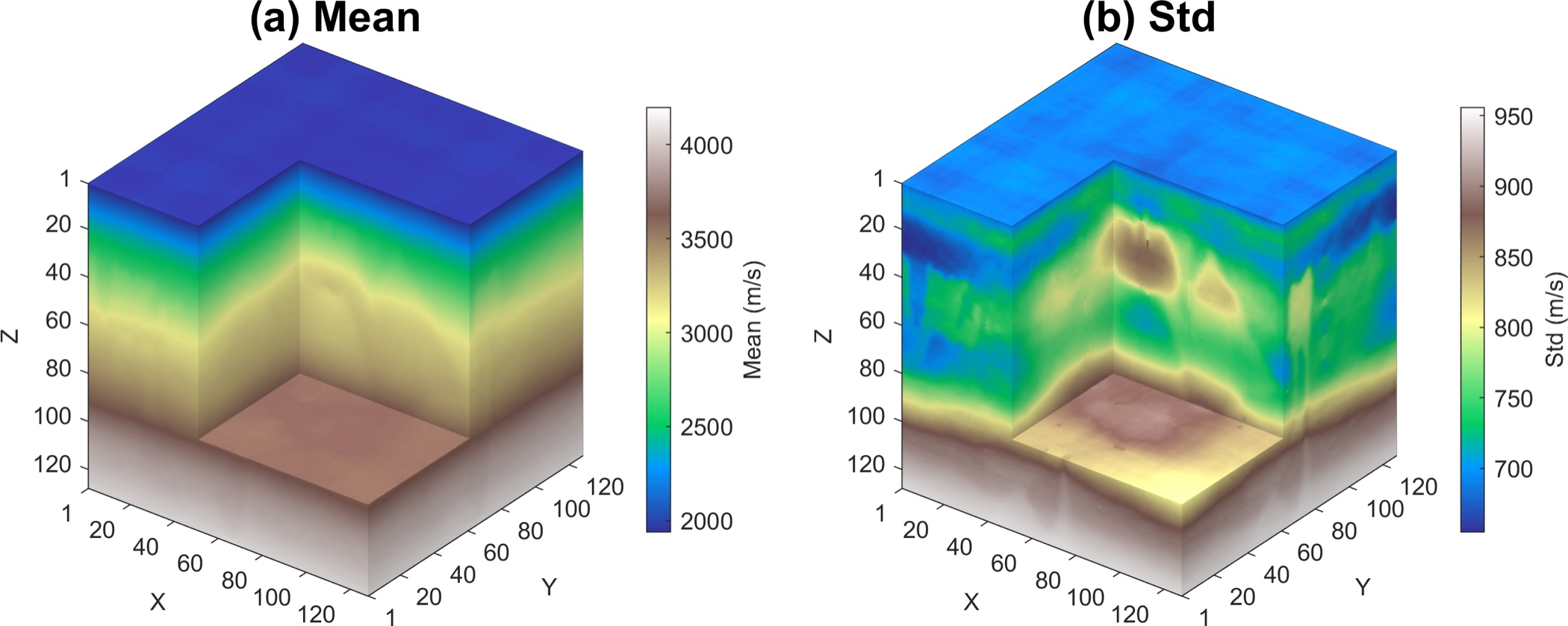}
\caption{Statistical properties of the full augmented training dataset of 22000 velocity models. (a) Mean velocity volume. (b) Standard deviation volume.}
\label{fig3}
\end{figure}

\begin{figure}[!htb] 
\centering
\includegraphics[width=1\textwidth]{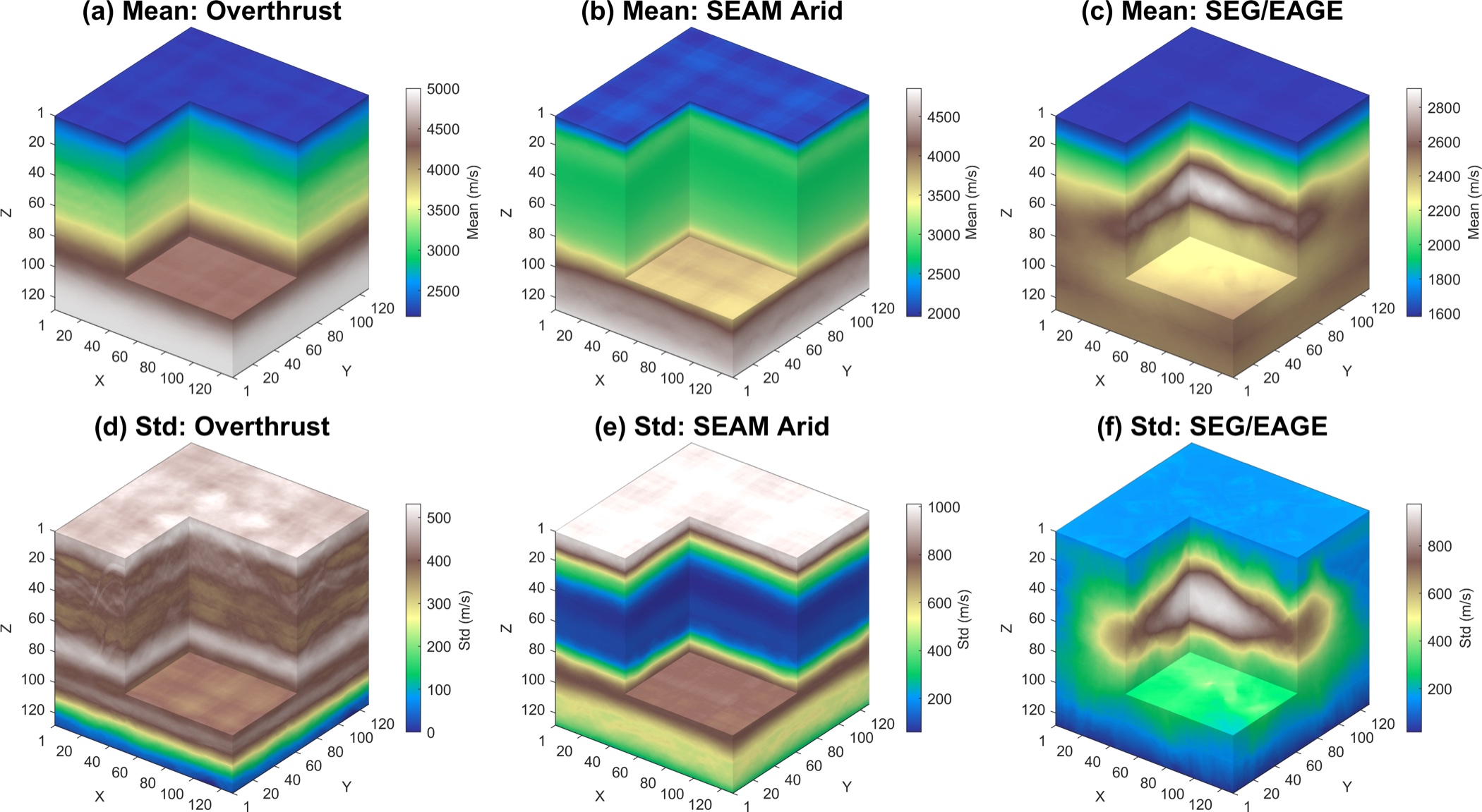}
\caption{Per-model statistics of the training samples derived from the three 3D industrial models used in subsequent experiments. The first row shows the mean velocity and the second row shows the standard deviation, from left to right corresponding to Overthrust, SEAM Arid, and SEG/EAGE.}
\label{fig4}
\end{figure}

\subsection{Training data and configuration}

The training dataset is built from both 2D and 3D industrial velocity models. The 2D component includes BP1994, BP2004, Hess, Otway, and Sigsbee models. Each model is resized to $128 \times 128$ grid points and used to build 500 3D velocity samples, as described below, giving 2500 samples. To include more complex 3D structural patterns, we further extract 2D slices from three 3D industrial models, namely Overthrust, SEAM Arid, and SEG/EAGE. These slices are also resized to $128 \times 128$, and each source model contributes 1000 samples. Before augmentation, the full dataset contains 5500 3D velocity models of size $128 \times 128 \times 128$.

Each 3D velocity model is generated from a 2D base model using the undulation-based extension. For each sample, two random displacement curves are independently generated along the $y$ direction. These curves are composed of a B-spline long-wavelength component and a multi-sine short-wavelength component, which control the lateral variations in the $z$ and $x$ directions, respectively. The 2D base model is then resampled at each $y$ slice according to the generated displacements, resulting in a 3D velocity volume with spatially varying lateral structures. The structural attribute used for conditioning is a synthetic seismic image obtained by convolving the vertical reflectivity of each velocity model with a Ricker wavelet. The dominant frequency of the wavelet is 60~Hz, and the time sampling interval is $\Delta t = 10^{-2}$~s. Although more realistic structural constraints could be obtained from migrated seismic data, this convolution image provides a simple and controlled proxy for proof-of-concept validation. Representative training samples from each source model are shown in Figure~\ref{fig2}.

To increase the diversity of the training data, each generated 3D model is augmented by rotating it $90^\circ$, $180^\circ$, and $270^\circ$ about the vertical axis. This augmentation expands the dataset by a factor of four, resulting in 22000 velocity models in total. Figure~\ref{fig3} shows the mean and standard deviation volumes computed from the full augmented dataset. Since the following in-distribution experiments are conducted on the Overthrust, SEAM Arid, and SEG/EAGE models, we also compute the corresponding per-model statistics using all training samples derived from each of these three sources. The results are shown in Figure~\ref{fig4}, where the first row presents the mean velocity and the second row presents the standard deviation. These statistics provide a useful summary of the prior information encoded in the training distribution.

The proposed network is trained for 200000 iterations with a batch size of 24 using the AdamW optimizer. The learning rate is fixed at $10^{-4}$. To improve the stability of sampling, we maintain an exponential moving average of the model weights with a decay rate of 0.999. Training is performed on a single NVIDIA A100 GPU and takes approximately 36 hours. The classifier-free guidance dropout probability is set to 5\% for both the structural attribute $s$ and the well constraint $w$, and the two dropout operations are applied independently.

For comparison, we also train a 3D convolutional benchmark using the same depth-progressive strategy. In this benchmark, the 2D U-Net is replaced by a 3D U-Net, and the depth dimension is treated as a spatial dimension rather than as the channel dimension, like in our previous work for 2D (discussed in Section \ref{sec:review}). Due to the much higher memory consumption of 3D convolutions, the batch size is reduced to 5. In addition, the number of feature channels at each U-Net stage is reduced by half relative to the proposed 2D network. Since the 3D benchmark requires more training time, it is trained for 300000 iterations, taking approximately 132 hours on the same GPU. The training dataset and other training settings are kept the same as those used for the proposed method.

In the following experiments, both the proposed method and the 3D benchmark are evaluated under four conditioning scenarios: unconditional, well-only, image-only, and well-plus-image. The unconditional case uses only the shallow velocity prior, while the well-only and image-only includes the shallow velocity prior as well. The well-plus-image case uses both constraints simultaneously. For stochastic inference, 50 samples are generated for each setting. To accelerate sampling, we adopt denoising diffusion implicit model~\citep{song2020denoising} with 10 denoising steps in place of the full 1000-step denoising diffusion probabilistic model chain, which reduces inference time substantially while producing results of comparable quality. The ensemble mean is used as the final prediction for error calculation, while the ensemble standard deviation is used to assess the variability of the sampled solutions.

\begin{figure}[!htb] 
\centering
\includegraphics[width=1\textwidth]{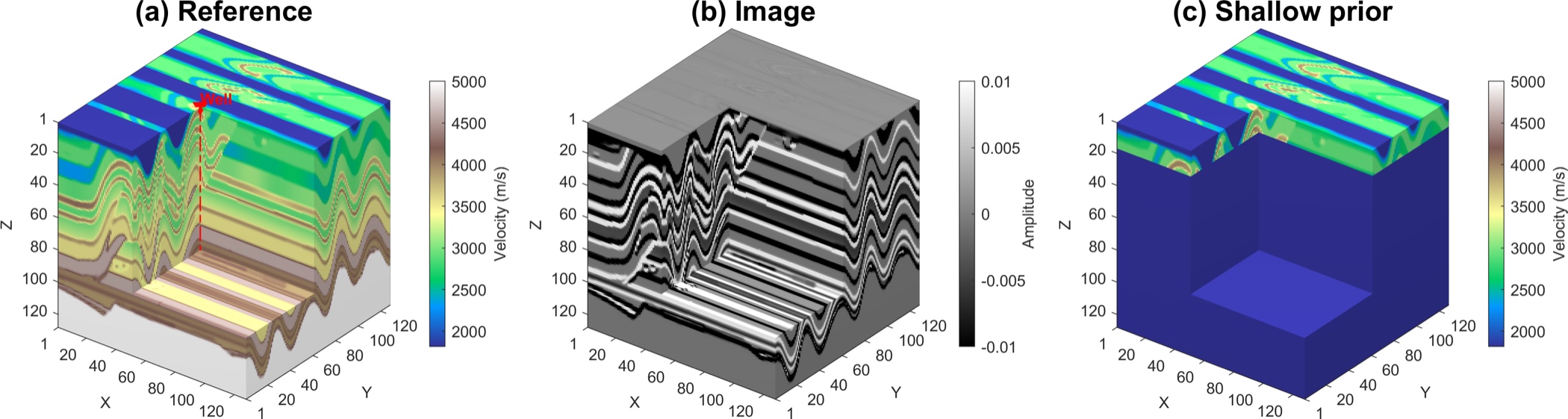}
\caption{Test configuration for the Overthrust model. (a) Reference 3D velocity model, with the red dashed line indicating the location of the single well constraint. (b) Convolution image used as the structural attribute. (c) Shallow velocity prior formed from the top 16 grid points of the reference model.}
\label{fig5}
\end{figure}

\begin{figure}[!htb] 
\centering
\includegraphics[width=1\textwidth]{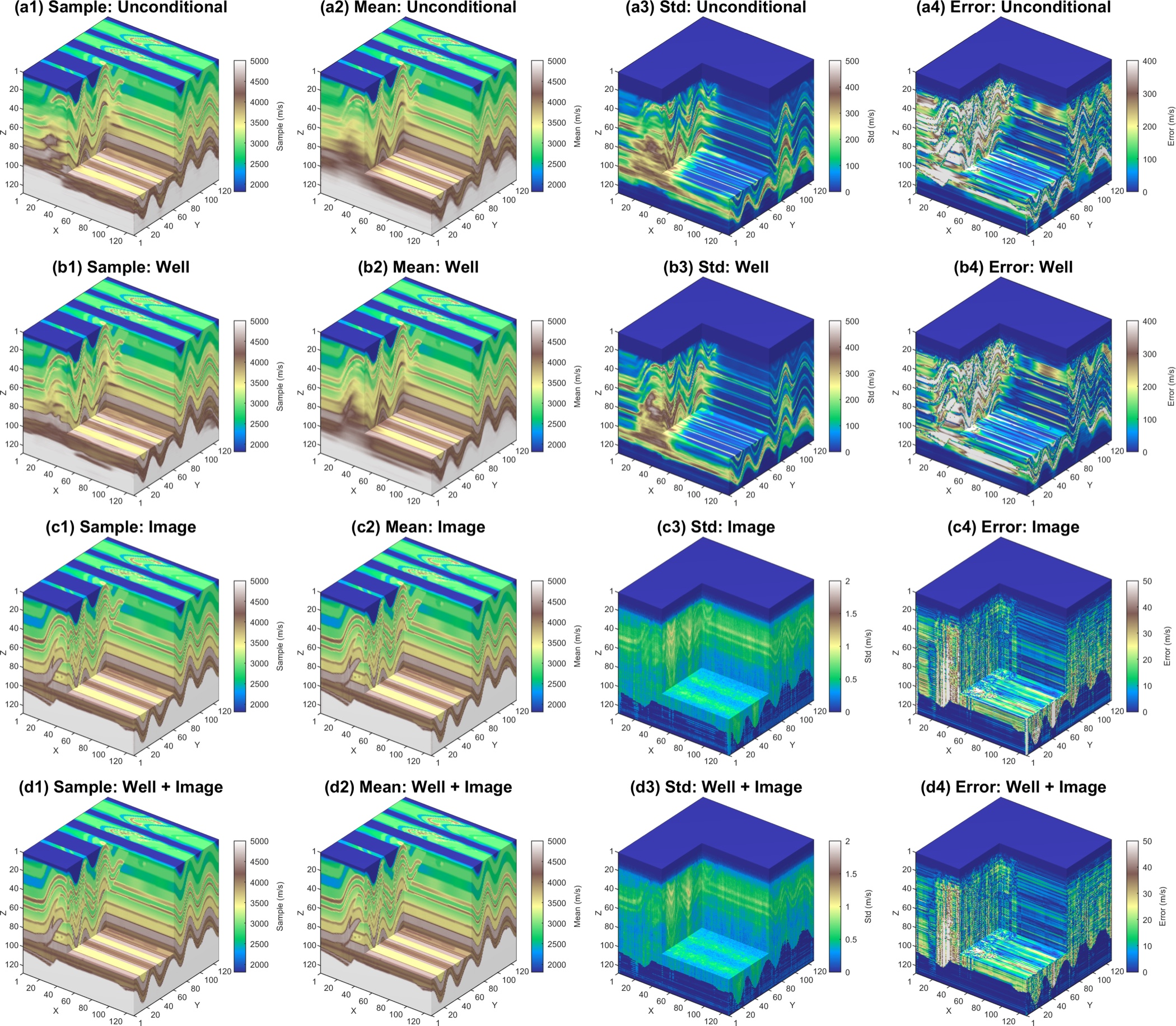}
\caption{3D VMB results on the Overthrust model obtained by the proposed method under four conditioning scenarios. Rows (a)-(d) correspond to unconditional, well-only, image-only, and well-plus-image conditioning, respectively. Columns 1-4 show a representative sample, the ensemble mean, the ensemble standard deviation, and the pointwise absolute error between the ensemble mean and reference model (Figure \ref{fig5}a).}
\label{fig6}
\end{figure}

\begin{figure}[!htb] 
\centering
\includegraphics[width=1\textwidth]{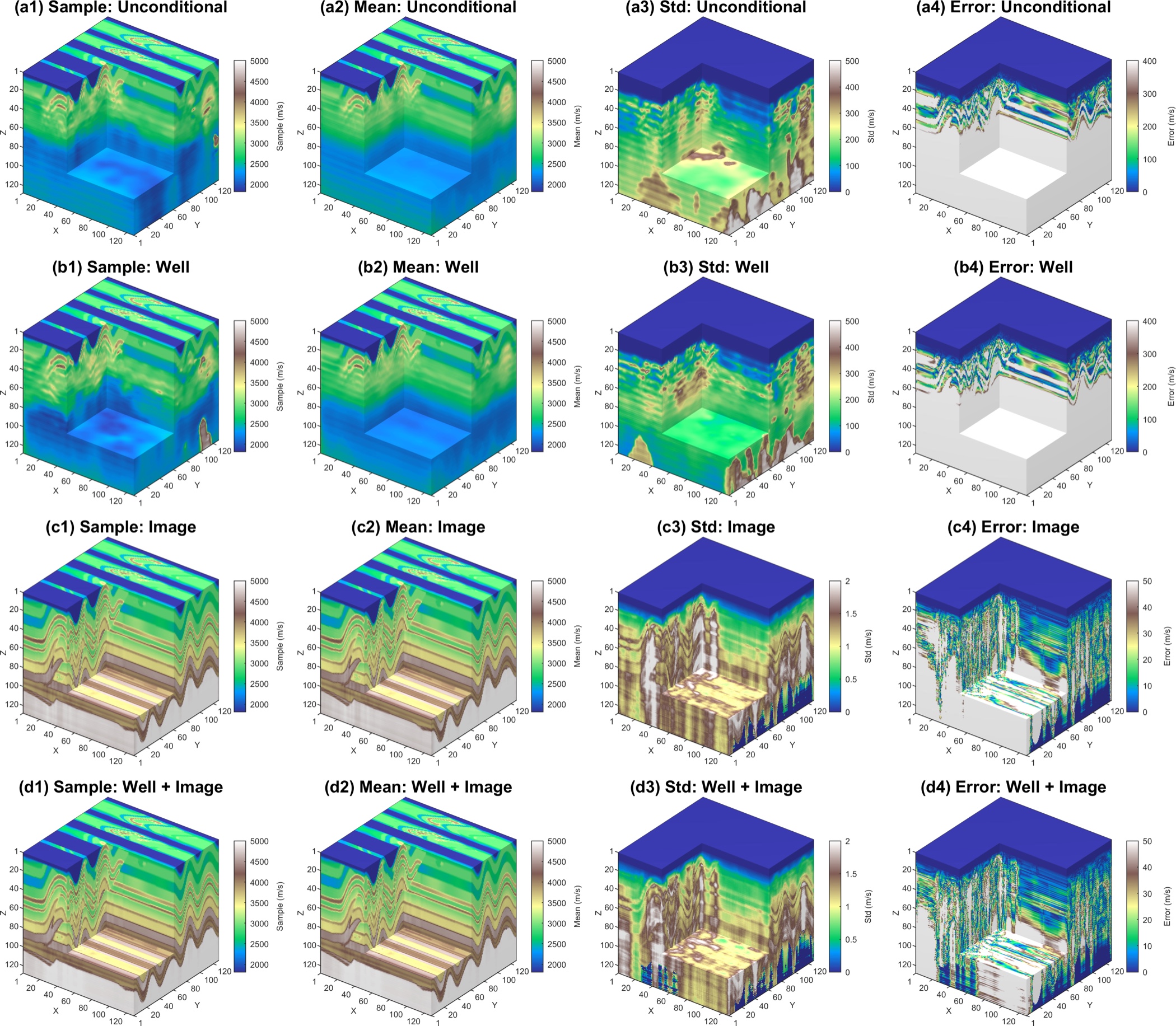}
\caption{3D VMB results on the Overthrust model obtained by the 3D network benchmark method under the same four conditioning scenarios as in Figure~\ref{fig6}.}
\label{fig7}
\end{figure}

\subsection{In-distribution test: Overthrust model}

We first consider the Overthrust model as an in-distribution test. For the in-distribution experiments, the test models are generated from unseen 2D slices extracted from the original Overthrust, SEAM Arid, and SEG/EAGE volumes. These slices are resized to $128 \times 128$ and passed through the same undulation-based 3D generation procedure used for training. Therefore, the test samples follow the same generation process as the training data, while no test slice is used during training.

Figure~\ref{fig5} shows the test configuration for the Overthrust example. The reference velocity model contains a characteristic fault system with moderate lateral velocity variations. The convolution image captures the main stratigraphic reflectors and is used as the structural constraint. The shallow prior is formed from the top 16 depth samples of the reference model, while the remaining depth samples are filled with a constant background velocity. A single well is placed at the lateral position $x=51$ and $y=61$, as indicated by the red dashed line in Figure~\ref{fig5}a, and provides point velocity measurements along the full depth column.

Figure~\ref{fig6} presents the 3D VMB results of the proposed method under four conditioning scenarios: unconditional, well-only, image-only, and well-plus-image. For each scenario, we show a representative sample, the ensemble mean, the ensemble standard deviation, and the pointwise absolute error. In the unconditional case (row a), the proposed method demonstrates its ability to propagate the shallow prior information progressively into the deeper regions: as seen in the $y$-$z$ cross-sections, the predicted velocity structure and magnitude agree reasonably well with the reference model throughout the depth column, confirming that the recursive depth-progressive inference effectively transfers shallow constraints downward even in the absence of additional conditioning. The primary limitation in this case lies in the fault zones, where complex structural patterns are not well recovered due to the lack of structural guidance, leading to elevated errors in those regions. Incorporating the well constraint (row b) provides modest improvement in the absolute velocity level but does not substantially reduce uncertainty, as a single well offers limited spatial coverage over the full 3D volume. The image constraint (row c) produces a markedly different outcome: the structural information guides the diffusion model to recover the layered reflectors with considerably higher fidelity, and the ensemble standard deviation drops by more than two orders of magnitude to below 2~m/s, indicating that the structural attribute effectively collapses the generative uncertainty. The combined well-plus-image conditioning (row d) yields the most accurate predictions, with further reduction in error compared to the image-only case, demonstrating that well and structural constraints provide complementary information that jointly constrain the solution.

Figure~\ref{fig7} shows the corresponding results for the 3D network benchmark under the same four conditioning scenarios. In the unconditional and well-only cases (rows a and b), the benchmark produces significantly larger errors in the deeper portion of the model compared to the proposed method, with error values noticeably elevated throughout the lower half of the volume. Even under image and well-plus-image conditioning (rows c and d), visible errors persist at depth, in contrast to the substantially cleaner predictions of the proposed method. A likely contributing factor is the inherent difficulty of optimizing a 3D convolutional network: with a considerably larger number of trainable parameters, the 3D network requires more iterations to converge and is more susceptible to underfitting within a fixed training budget. This optimization challenge manifests most clearly at greater depths, where the recursive propagation of errors is most pronounced and the network has the least capacity to recover.

\begin{figure}[!htb] 
\centering
\includegraphics[width=1\textwidth]{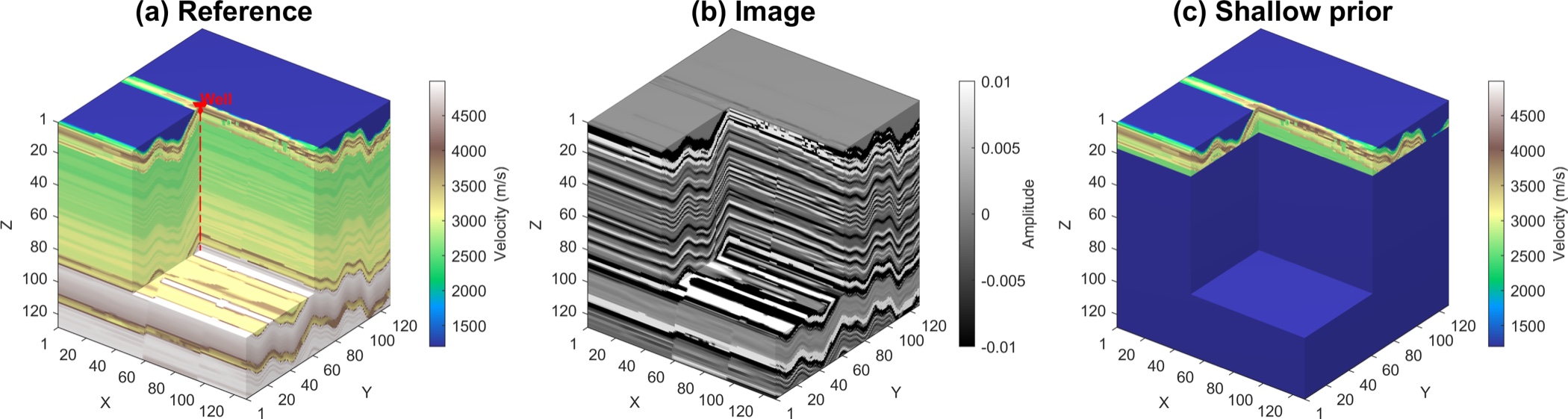}
\caption{Test configuration for the SEAM Arid model. (a) Reference 3D velocity model, with the red dashed line indicating the well location. (b) Convolution image used as the structural constraint. (c) Shallow velocity prior.}
\label{fig8}
\end{figure}

\begin{figure}[!htb] 
\centering
\includegraphics[width=1\textwidth]{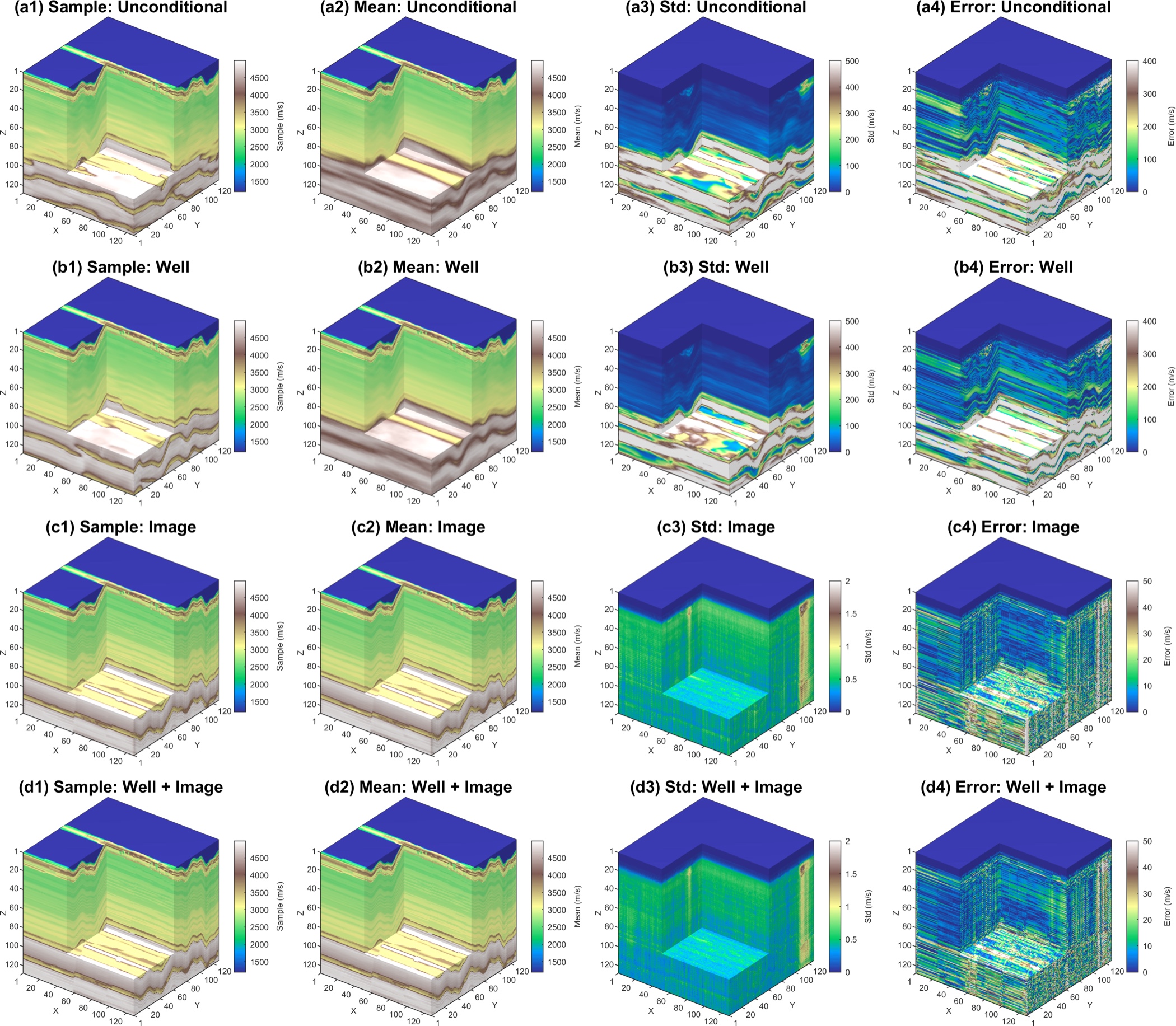}
\caption{3D VMB results on the SEAM Arid model obtained by the proposed method under four conditioning scenarios. Rows (a)-(d) correspond to unconditional, well-only, image-only, and well-plus-image conditioning, respectively. Columns 1-4 show a representative sample, the ensemble mean, the ensemble standard deviation, and the pointwise absolute error between the ensemble mean and reference model (Figure \ref{fig8}a).}
\label{fig9}
\end{figure}

\begin{figure}[!htb] 
\centering
\includegraphics[width=1\textwidth]{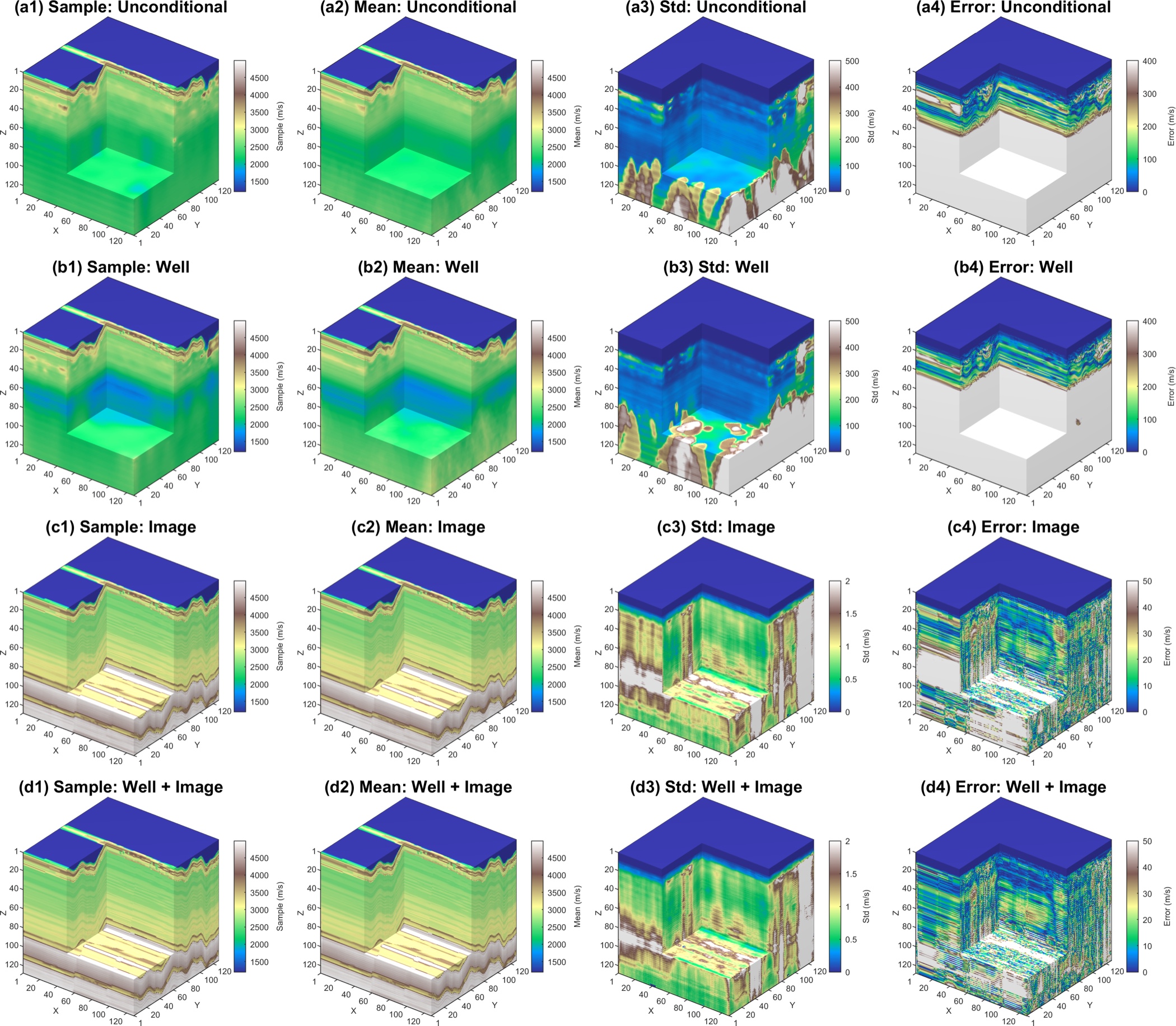}
\caption{3D VMB results on the SEAM Arid model obtained by the 3D network benchmark method under the same four conditioning scenarios as in Figure~\ref{fig9}.}
\label{fig10}
\end{figure}

\subsection{In-distribution test: SEAM Arid model}

The SEAM Arid model provides a more challenging in-distribution example. As shown in Figure~\ref{fig8}a, the model contains complex shallow features and a pronounced high-velocity carbonate anomaly in the lower part of the volume, with strong velocity contrasts against the surrounding sedimentary layers. The convolution image in Figure~\ref{fig8}b reflects this complexity, with high-amplitude events associated with the boundaries of the anomaly body. The shallow prior and well configuration follow the same setup as in the Overthrust experiment.

Figure~\ref{fig9} shows the results of the proposed method. In the unconditional case (row a), the depth-progressive inference successfully propagates the shallow velocity trend into deeper regions, recovering the broad velocity gradient. The general information controlling the behavior of the model deep is stored in the diffusion model, and this particular behavior is guided by the shallow features, which the network associates with an Arid model (considering the in distribution nature of the teat). However, the high-velocity anomaly body is not well resolved, and the ensemble standard deviation reaches up to approximately 500~m/s in that region, indicating that the complex deep structure cannot be reliably inferred from the shallow prior alone. The well constraint (row b) provides limited additional benefit in this case, as a single well offers insufficient spatial coverage to constrain the lateral extent of the anomaly. The image constraint (row c) proves critical for this model: the structural attribute directly reveals the geometry of the high-velocity body, enabling the diffusion model to recover its shape and velocity magnitude with high fidelity. The ensemble standard deviation drops dramatically to below 2~m/s, and the error map shows only minor residuals. The well-plus-image case (row d) achieves the lowest overall error, with the well further anchoring the absolute velocity level.

Figure~\ref{fig10} shows the benchmark results. Consistent with the Overthrust observations, the benchmark exhibits substantially larger errors at depth under unconditional and well-only conditioning. Under image conditioning (row c), the benchmark recovers the general shape of the high-velocity anomaly but leaves visible velocity errors at the anomaly boundaries and in the deeper layers below it. These residuals are more pronounced than in the proposed method, suggesting that the 3D network, despite its larger parameter count, struggles to accurately model the sharp velocity contrasts in this model within the given training budget.

\begin{figure}[!htb] 
\centering
\includegraphics[width=1\textwidth]{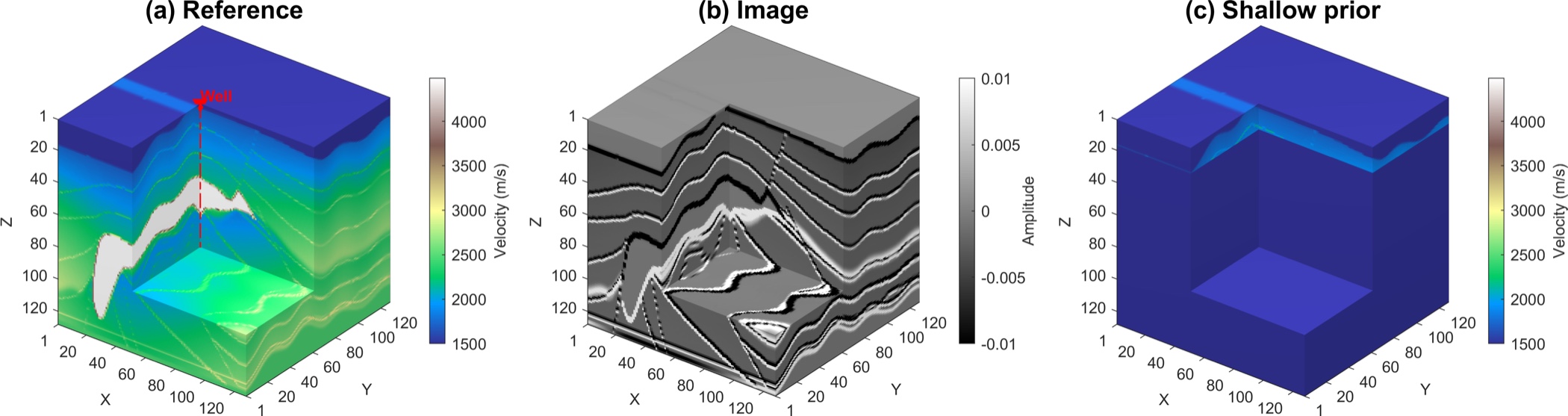}
\caption{Test configuration for the SEG/EAGE model. (a) Reference 3D velocity model, with the red dashed line indicating the well location. (b) Convolution image used as the structural constraint. (c) Shallow velocity prior.}
\label{fig11}
\end{figure}

\begin{figure}[!htb] 
\centering
\includegraphics[width=1\textwidth]{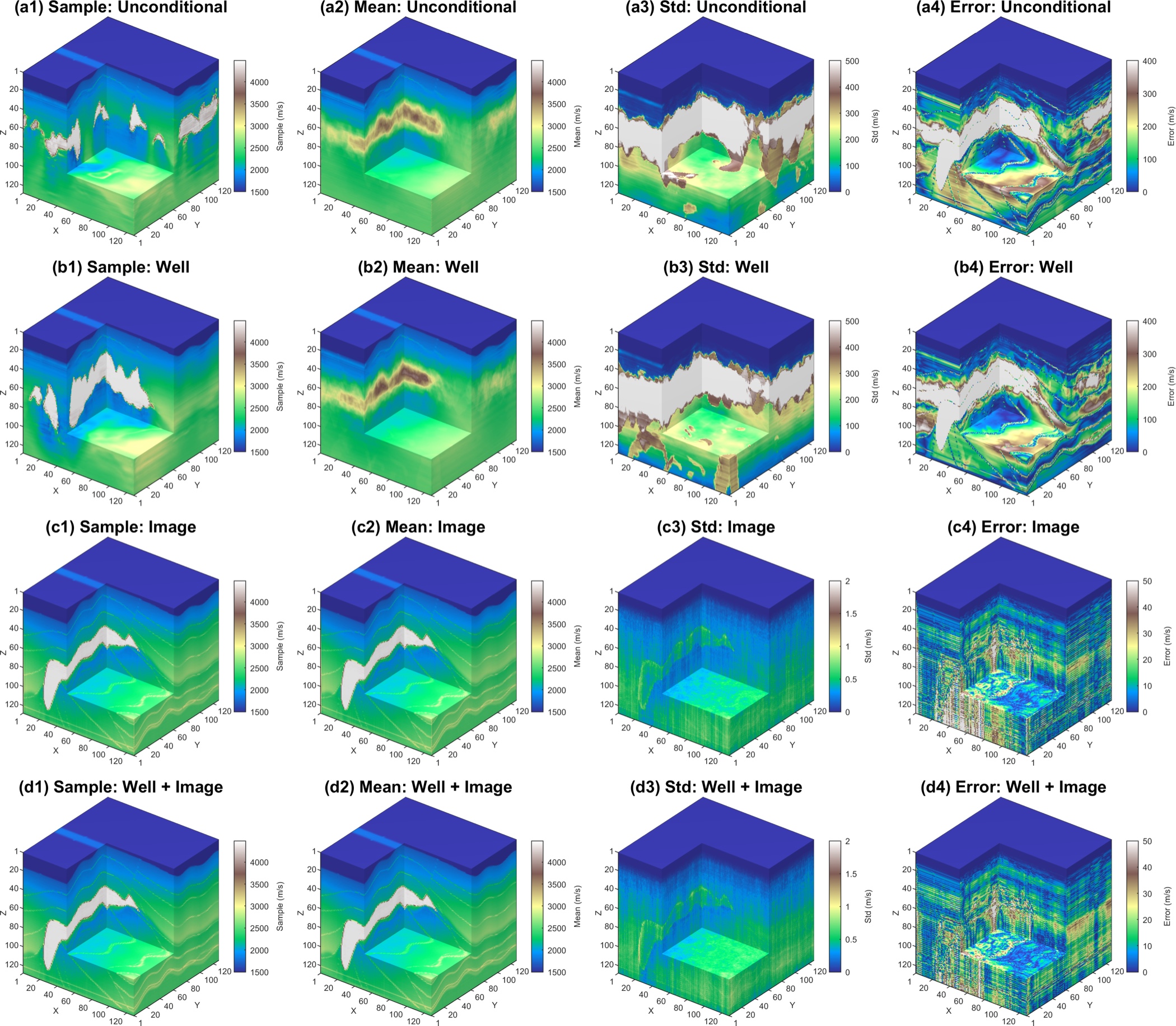}
\caption{3D VMB results on the SEG/EAGE model obtained by the proposed method under four conditioning scenarios. Rows (a)-(d) correspond to unconditional, well-only, image-only, and well-plus-image conditioning, respectively. Columns 1-4 show a representative sample, the ensemble mean, the ensemble standard deviation, and the pointwise absolute error between the ensemble mean and reference model (Figure \ref{fig11}a).}
\label{fig12}
\end{figure}

\begin{figure}[!htb] 
\centering
\includegraphics[width=1\textwidth]{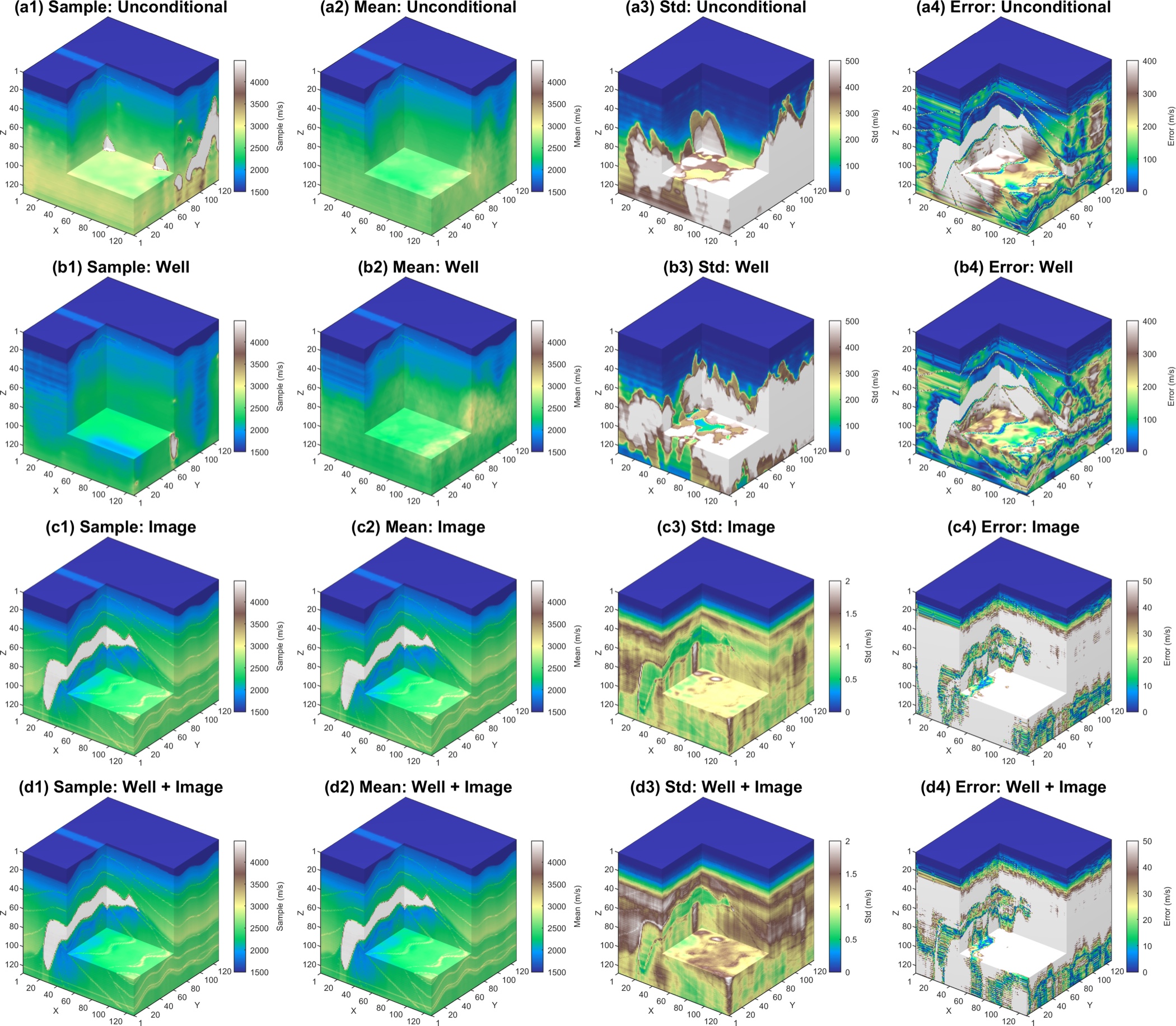}
\caption{3D VMB results on the SEG/EAGE model obtained by the 3D network benchmark method under the same four conditioning scenarios as in Figure~\ref{fig12}.}
\label{fig13}
\end{figure}

\subsection{In-distribution test: SEG/EAGE model}

The SEG/EAGE model is the most structurally complex of the three in-distribution examples. As shown in Figure~\ref{fig11}a, it contains a large salt body with irregular 3D geometry embedded in a layered sedimentary background. The sharp and spatially variable velocity contrast between the salt and the surrounding sediments makes this example a stringent test of the model's sensitivity to structural information. The convolution image in Figure~\ref{fig11}b captures the strong reflections around the salt body and the relatively transparent salt interior, thereby providing useful geometric guidance. In contrast, the shallow prior in Figure~\ref{fig11}c only covers the near-surface sedimentary layers and does not contain direct information about the salt body.

Figure~\ref{fig12} shows the results of the proposed method. In the unconditional and well-only cases (rows a and b), individual samples do not accurately recover the salt geometry. Nevertheless, the ensemble mean in both cases reveals a velocity anomaly at approximately the correct depth and lateral position of the salt body, while the standard deviation maps show elevated uncertainty concentrated at the depths where the salt body resides, reflecting the limited ability to localize the anomaly without structural guidance. Again, the prediction of the Salt is guided by the nature of the shallow part, which the network, thanks to the training samples, associate with Salt bodies. The well guidance helped build a more accurate salt body, at least around the well location. The image constraint (row c) resolves this issue: the salt geometry is clearly delineated in both the individual sample and the ensemble mean, with the irregular salt flanks reproduced with good fidelity. The standard deviation drops to below 2~m/s across most of the volume, with slightly elevated values near the more complex portions of the salt boundary. The well-plus-image result (row d) further reduces the pointwise error, particularly in the sedimentary layers surrounding the salt body.

The benchmark results in Figure~\ref{fig13} reveal a qualitatively different behavior. In the unconditional and well-only cases, the ensemble mean does not exhibit any velocity anomaly near the salt body location, in contrast to the proposed method. This suggests that the 3D network has not effectively learned the prior characteristics of the SEG/EAGE distribution from the training data, and the recursive inference fails to generate salt-bearing models even when guided by the shallow prior. Under image and well-plus-image conditioning, the benchmark recovers the broad outline of the salt body, but the uncertainty around the salt body is noticeably larger than that of the proposed method, and the pointwise error maps show more significant residuals both at the salt boundaries and in the surrounding sediments.

\begin{figure}[!htb] 
\centering
\includegraphics[width=1\textwidth]{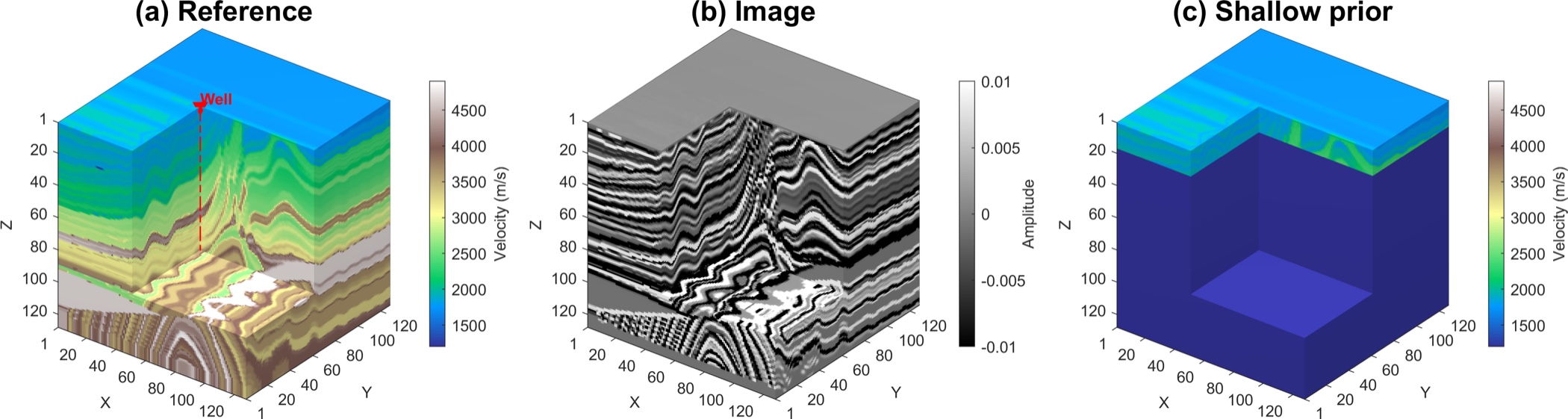}
\caption{Test configuration for the Marmousi model. (a) Reference 3D velocity model, with the red dashed line indicating the well location. (b) Convolution image used as the structural constraint. (c) Shallow velocity prior.}
\label{fig14}
\end{figure}

\begin{figure}[!htb] 
\centering
\includegraphics[width=1\textwidth]{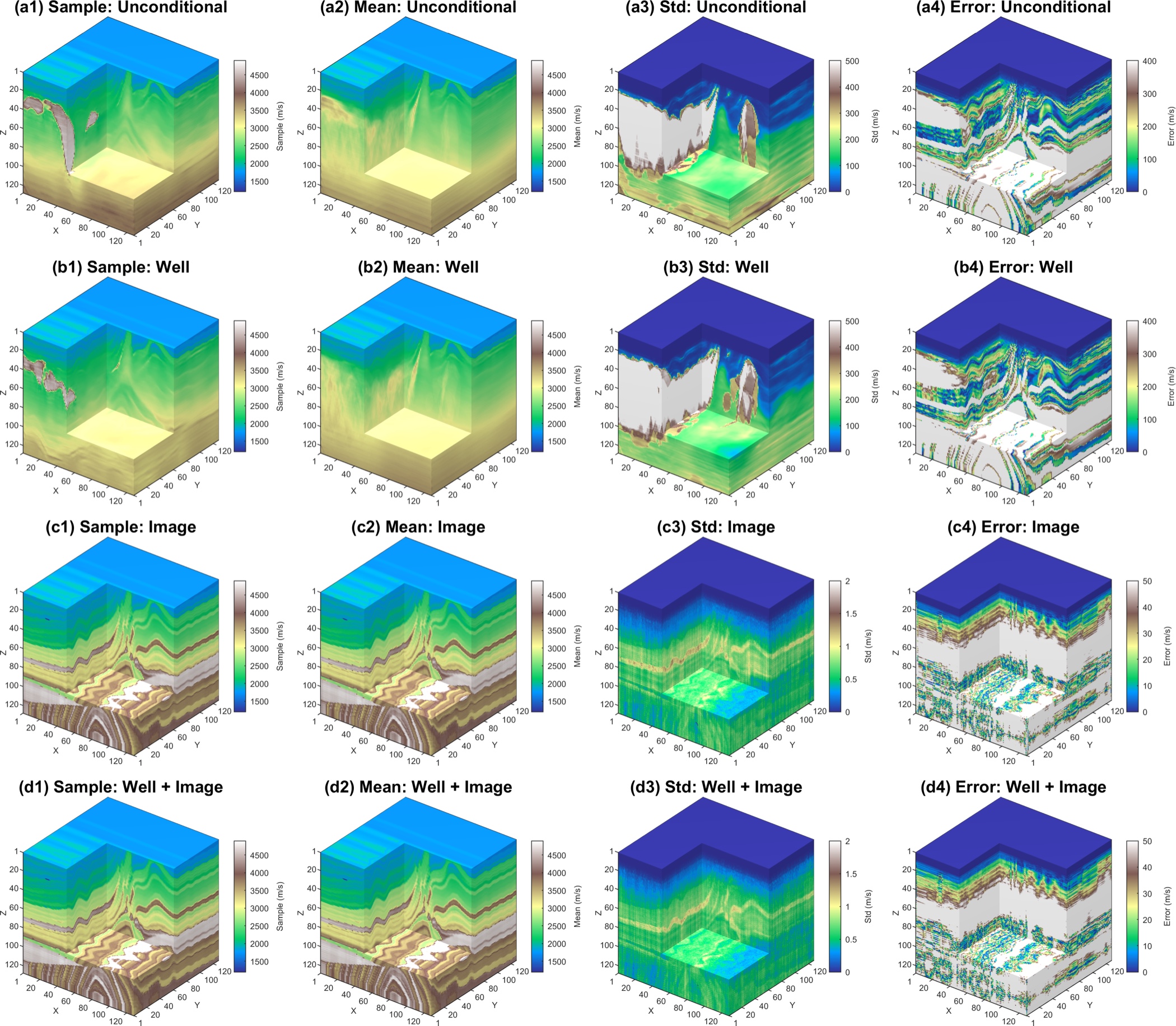}
\caption{3D VMB results on the Marmousi model obtained by the proposed method under four conditioning scenarios. Rows (a)-(d) correspond to unconditional, well-only, image-only, and well-plus-image conditioning, respectively. Columns 1-4 show a representative sample, the ensemble mean, the ensemble standard deviation, and the pointwise absolute error between the ensemble mean and reference model (Figure \ref{fig14}a).}
\label{fig15}
\end{figure}

\begin{figure}[!htb] 
\centering
\includegraphics[width=1\textwidth]{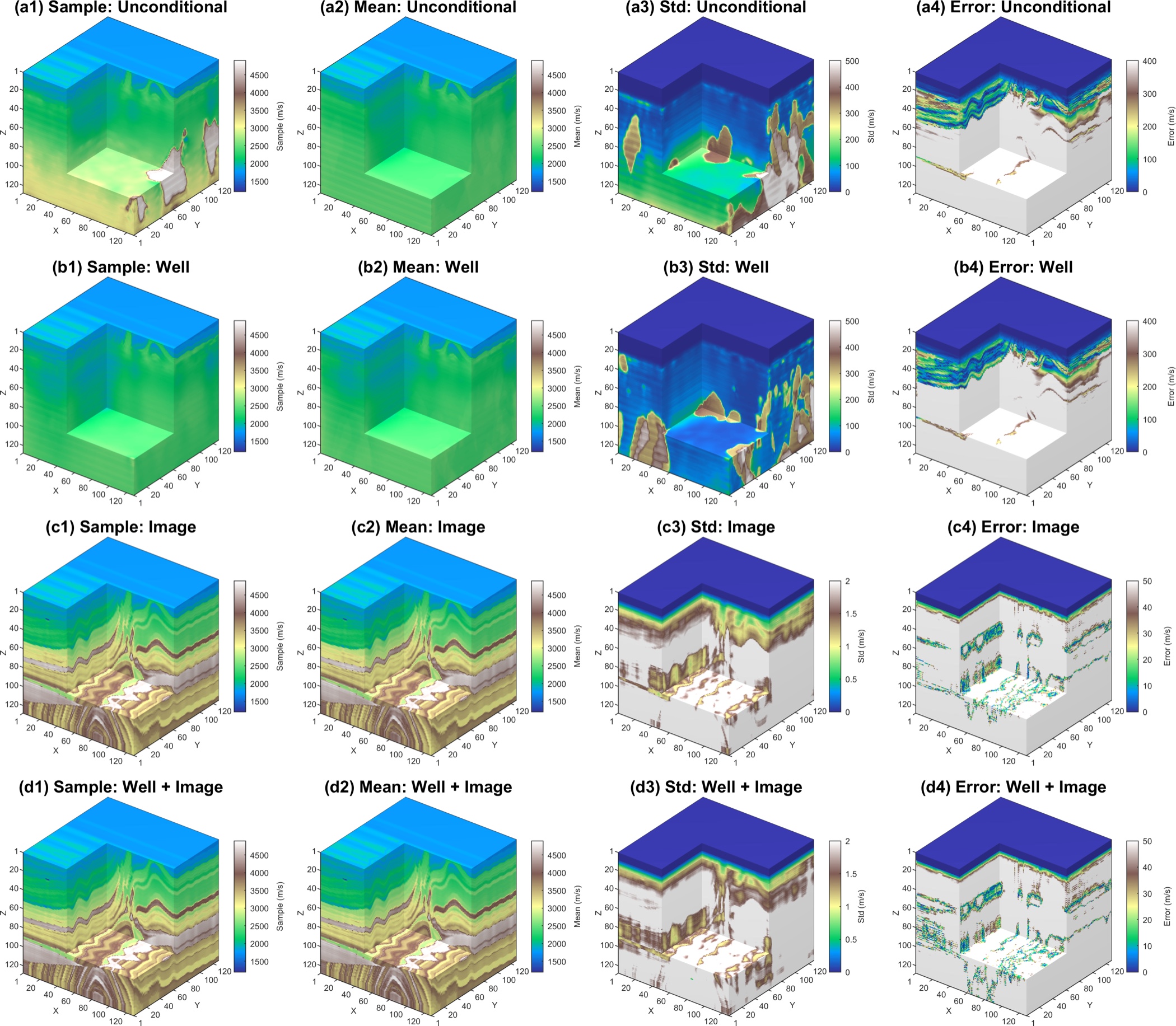}
\caption{3D VMB results on the Marmousi model obtained by the 3D network benchmark method under the same four conditioning scenarios as in Figure~\ref{fig15}.}
\label{fig16}
\end{figure}

\subsection{Out-of-distribution test: Marmousi model}

The Marmousi model serves as the out-of-distribution test case, just in case the solution deviated from our prior knowledge, evaluating the generalization ability of the proposed method beyond the training distribution. As shown in Figure~\ref{fig14}a, the model is characterized by complex thrust fault structures, strong lateral velocity gradients, and a high-velocity body in the deeper portion of the model, none of which are directly represented in the training data. The convolution image (Figure~\ref{fig14}b) reflects the structural complexity with densely packed, laterally discontinuous reflectors. The shallow prior (Figure~\ref{fig14}c) captures only the near-surface portion of the model.

Figure~\ref{fig15} shows the results of the proposed method. Without structural guidance (rows a and b), the predictions recover the broad vertical velocity trend but fail to resolve the lateral discontinuities and fault structures, with the ensemble standard deviation remaining elevated throughout the complex mid-to-deep portion of the model. These results are expected given the out-of-distribution nature of the test. However, a closer look reveals that network managed to extend some of the faults beyond he provided shallow patch, as the network learned from the training samples such features. Under image conditioning (row c), the structural complexity of the Marmousi model is substantially better captured: the layered structure and the lateral velocity variations are visible in both the individual sample and the ensemble mean, and the standard deviation drops markedly compared to the unconstrained cases. The well-plus-image result (row d) achieves the lowest pointwise error among the four scenarios, though some residuals persist around the fault zones where the structural complexity exceeds what the model has encountered during training. These results demonstrate a reasonable degree of generalization to out-of-distribution geology, particularly when structural constraints are available.

Figure~\ref{fig16} shows the benchmark results on the same model. The unconditional and well-only cases produce smoother predictions with less structural detail than the proposed method, and the error maps show larger residuals throughout the model volume. Under image conditioning, the benchmark recovers the general reflector pattern but produces noticeably larger errors in the fault zones and around the high-velocity body at depth. The well-plus-image case improves somewhat but the overall error level remains higher than that of the proposed method, suggesting that the 3D network generalizes less effectively to out-of-distribution geology under the same training conditions.

\begin{table*}[t]
\centering
\caption{Quantitative comparison of MAE (m/s) and SSIM for the proposed method and the 3D benchmark under four conditioning scenarios across all test models. The best result in each row is shown in \textbf{bold}.}
\label{tab1}
\begin{tabular}{llcccccccc}
\toprule
& & \multicolumn{4}{c}{Proposed Method} & \multicolumn{4}{c}{3D Benchmark} \\
\cmidrule(lr){3-6} \cmidrule(lr){7-10}
Model & Metric & Uncond. & Well & Image & Well+Image & Uncond. & Well & Image & Well+Image \\
\midrule
\multirow{2}{*}{Overthrust}
& MAE   & \textbf{129.83} & \textbf{127.80} & \textbf{6.89} & \textbf{6.88} & 1239.42 & 1261.15 & 24.90 & 20.53 \\
& SSIM  & \textbf{0.757}  & \textbf{0.761}  & \textbf{0.999} & \textbf{0.999} & 0.410 & 0.410 & 0.993 & 0.996 \\
\midrule
\multirow{2}{*}{SEAM Arid}
& MAE   & \textbf{227.49} & \textbf{247.79} & \textbf{11.56} & \textbf{11.33} & 932.69 & 958.78 & 28.58 & 24.56 \\
& SSIM  & \textbf{0.637}  & \textbf{0.621}  & \textbf{0.993} & \textbf{0.994} & 0.615 & 0.605 & 0.987 & 0.989 \\
\midrule
\multirow{2}{*}{SEG/EAGE}
& MAE   & \textbf{247.53} & \textbf{253.35} & \textbf{11.38} & \textbf{11.01} & 271.83 & 268.69 & 58.43 & 56.72 \\
& SSIM  & 0.687           & 0.688  & \textbf{0.995} & \textbf{0.996} & \textbf{0.731} & \textbf{0.713} & 0.983 & 0.987 \\
\midrule
\multirow{2}{*}{Marmousi}
& MAE   & \textbf{464.57} & \textbf{425.96} & \textbf{49.22} & \textbf{48.45} & 900.06 & 931.91 & 143.44 & 135.81 \\
& SSIM  & \textbf{0.450}  & \textbf{0.446}  & \textbf{0.994} & \textbf{0.994} & 0.442 & 0.440 & 0.969 & 0.973 \\
\bottomrule
\end{tabular}
\end{table*}

\subsection{Quantitative comparison}

Having qualitatively examined the results visually across all four test models, we now turn to a quantitative summary. Table~\ref{tab1} provides a quantitative summary of the visual comparisons discussed above. The overall trends are clear and consistent across the four test models. First, the structural image is the most influential conditioning information. Moving from the unconditional or well-only cases to the image-conditioned cases leads to a pronounced reduction in MAE and a substantial increase in SSIM for both methods. This confirms that the structural image not only improves the geometric consistency of the reconstructed velocity models, but also helps constrain the velocity magnitude. In contrast, the single-well constraint alone brings only limited improvement, and in some cases the metrics change only slightly. This behavior is expected because one vertical well provides accurate but spatially limited velocity information, whereas the image constraint contains broader structural guidance over the full model volume. When the well is combined with the image constraint, the MAE is further reduced in most cases, while the SSIM improvement is relatively small. This suggests that the well mainly refines the absolute velocity level once the dominant structural pattern has already been constrained by the image.

The comparison between in-distribution and out-of-distribution tests further illustrates the role of the training distribution. For the three in-distribution models, the proposed method achieves relatively low errors, especially when image conditioning is available. The Marmousi model, by contrast, shows larger errors under the same conditioning scenarios, reflecting the difficulty of applying the trained model to a geological style not represented in the training data. Nevertheless, the use of the structural image still leads to a large improvement on Marmousi, indicating that structural conditioning helps mitigate, although not completely remove, the effect of distributional shift.

Compared with the 3D benchmark, the proposed method achieves lower MAE in all test cases and conditioning scenarios. This supports the effectiveness of the depth-as-channel formulation for 3D VMB. The SSIM values show the same trend in most cases, especially when image conditioning is used. The only exceptions occur for the SEG/EAGE model under the unconditional and well-only settings, where the 3D benchmark gives slightly higher SSIM values. However, its MAE remains higher in both cases, indicating that the slightly improved structural similarity does not correspond to a more accurate velocity reconstruction. Once the structural image is introduced, the proposed method outperforms the 3D benchmark in both MAE and SSIM. These quantitative results confirm that the proposed 2D diffusion model provides a more accurate and stable alternative to the direct 3D convolutional benchmark, while benefiting most strongly from structural conditioning.
\section{\textbf{Discussion}}\label{sec:discussion}

The numerical experiments presented in the previous section demonstrate the effectiveness of the proposed framework across a range of geological settings and conditioning scenarios. In this section, we illustrate several aspects of the method in greater depth. We first discuss the respective roles of the training distribution prior and the shallow velocity prior, and how they interact within the depth-progressive inference framework. We then present an ablation study on the depth-attenuated Gaussian blending scheme, followed by a comparison of computational efficiency between the proposed method and the 3D benchmark. Finally, we discuss the current limitations of the method and outline directions for future work.

\subsection{Role of training distribution prior and shallow prior}

The proposed framework relies on two complementary sources of prior information that work in concert throughout the depth-progressive inference: the statistical prior learned from the training distribution and the shallow velocity prior provided at inference time. The shallow prior functions as a form of geological class guidance: once the shallow velocity is given, it directs the diffusion model to identify the most compatible portion of the learned velocity distribution, and then uses that information as a starting point to progressively construct the velocity model toward greater depths.

This guidance mechanism can be illustrated through the three in-distribution test cases. On the Overthrust model, the shallow prior carries both the velocity magnitude characteristics and partial structural information of the thrust fault system. This is sufficient to guide the diffusion model, even under unconditional generation, to produce velocity predictions that broadly reproduce the depth trend and the approximate fault geometry throughout the model. On the SEAM Arid model, the shallow prior contains the fine-scale velocity structure characteristic of the near-surface sedimentary sequence of that model, which the diffusion model recognizes and uses to generate velocity volumes with features consistent with the SEAM Arid geological style. On the SEG/EAGE model, the shallow prior guides the model to generate a salt-like anomaly at the appropriate depth, as visible in the single sample of Figure~\ref{fig12}a. The ensemble mean in that case shows a central velocity anomaly, and if one compares this with the mean of the SEG/EAGE training samples in Figure~\ref{fig4}c, the anomaly appears at a matching depth range, reflecting how the training distribution shapes the expected velocity structure at each depth level.

Beyond guiding the structural character of the predictions, the training distribution also influences the spatial pattern of velocity model building (VMB) uncertainty. In general, uncertainty tends to increase with depth as the recursive inference propagates errors downward. However, the Overthrust unconditional uncertainty map in Figure~\ref{fig6}a3 shows an interesting exception: the uncertainty drops to near zero in the deepest portion of the model rather than continuing to increase. This behavior is explained by the training data statistics shown in Figure~\ref{fig4}d, where the standard deviation of the Overthrust training samples is very small at that depth, indicating that the velocity in this region is nearly constant across all training examples. The diffusion model therefore predicts this depth with high confidence regardless of the specific input, producing a markedly reduced uncertainty. This example illustrates that the VMB uncertainty at any given depth reflects not only the propagated error from the recursive inference but also the intrinsic variability of the training distribution at that location.

When structural constraints are absent, however, the two priors alone cannot resolve lateral ambiguity. As seen in the SEG/EAGE unconditional results, while the ensemble mean correctly identifies the approximate depth of the salt body, the individual samples place the anomaly at varying lateral positions, leading to elevated uncertainty at those depths. This lateral ambiguity is precisely what the structural image constraint resolves: by providing explicit geometric guidance, it collapses the uncertainty and allows the model to recover the correct three-dimensional structure. The well constraint further anchors the absolute velocity level at a specific lateral location, complementing the structural guidance. 

The interplay between these three sources of information, the training prior, the shallow prior, and the explicit geophysical constraints, is what gives the proposed framework its flexibility across a wide range of conditioning scenarios. These observations are consistent with those reported in our previous 2D depth-progressive VMB work \citep{chengdiffvmbpart1, chengdiffvmbpart2}, where the shallow prior was similarly found to act as a geological class indicator that guides the diffusion model toward geologically plausible predictions, and where the structural image constraint was likewise shown to be the dominant factor in resolving structural ambiguity. The extension to 3D does not alter these fundamental behaviors. Rather, the depth-as-channel formulation preserves them while enabling volumetric VMB at a fraction of the cost of a fully 3D approach.

\begin{figure}[!htb] 
\centering
\includegraphics[width=1\textwidth]{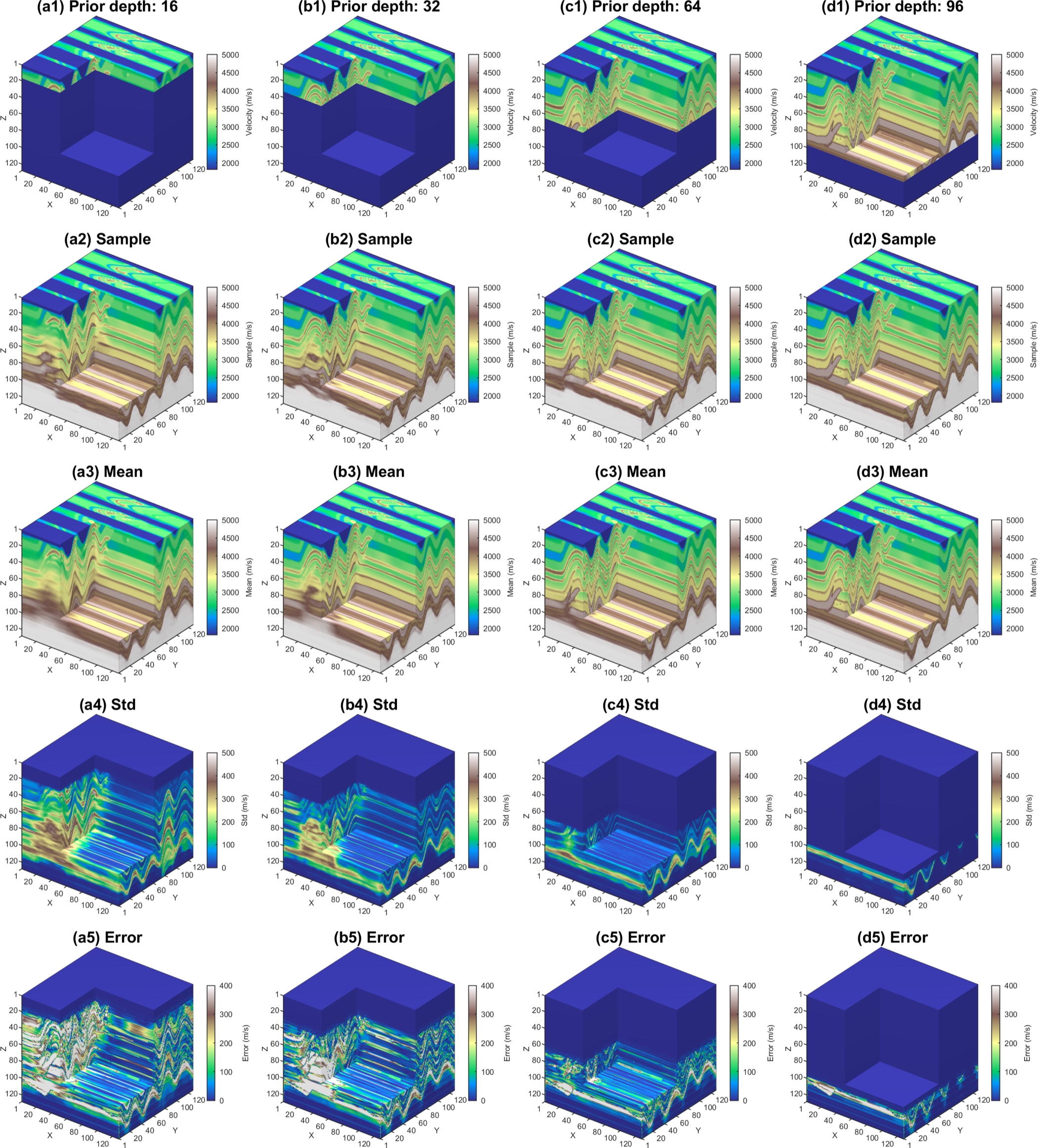}
\caption{Effect of shallow prior extent on unconditional 3D VMB results for the Overthrust model. Columns (a)-(d) correspond to shallow prior depths of 16, 32, 64, and 96 grid points, respectively. Rows 1-5 show the shallow prior, a representative single sample, the ensemble mean over 50 samples, the standard deviation, and the pointwise absolute error between the ensemble mean and reference model (Figure \ref{fig5}a).}
\label{fig17}
\end{figure}

\subsection{Effect of shallow prior extent on velocity model building}

In our previous 2D depth-progressive VMB work, we observed that a more informative shallow prior consistently leads to improved VMB accuracy. Here we verify whether this finding extends to the 3D setting. To isolate the contribution of the shallow prior from other conditioning signals, we conduct this analysis under unconditional generation on the Overthrust model, where the shallow prior is the only source of guidance. Figure~\ref{fig17} shows the results for four shallow prior extents of 16, 32, 64, and 96 grid points from the surface. Each column corresponds to one extent, and the rows show the shallow prior, a representative single sample, the ensemble mean over 50 samples, the standard deviation, and the pointwise absolute error.

The results confirm that the conclusion from the 2D setting carries over directly to 3D. With only 16 grid points of shallow prior, the individual samples and the ensemble mean recover the broad layered velocity trend but fail to reproduce the thrust fault structures in the deeper portion of the model, and the standard deviation map shows elevated uncertainty throughout the unconstrained depth range, particularly in the fault zones. As the prior extent increases to 32 grid points, the fault geometry begins to emerge in both the samples and the mean, though some ambiguity remains in the deeper layers, and the uncertainty is partially reduced. With 64 grid points, the layered structure and fault patterns are substantially better defined, and the standard deviation decreases markedly. With 96 grid points of prior, the thrust fault geometry is clearly recovered in both individual samples and the ensemble mean, closely matching the reference model (Figure \ref{fig5}a), and the uncertainty is reduced to a very low level throughout the remaining unpredicted depth range. The error maps confirm this trend quantitatively, with the pointwise absolute error decreasing progressively as the shallow prior extent grows and the errors becoming increasingly confined to the deepest and most structurally complex regions. These results demonstrate that a more extensive and reliable shallow prior consistently enhances 3D VMB accuracy, reduces generative uncertainty, and strengthens the model's ability to recover complex structural features, in full agreement with the observations from our prior 2D work.

Actually, this test reveals a more interesting behavior. That is, the deep patches extract information from the shallow regardless of depth. It implies that the mechanism or importance of the shallow velocity patch does not stop at the first prediction or just as a class, but that prior propagates to the deep, and considering how well surface recorded data constrain the shallow part of the Earth, that data source of information will also propagate to the deep.

\begin{table}[t]
\centering
\caption{Computational comparison between the proposed method and the 3D benchmark under equivalent network configurations. Training cost is measured per 100 iterations at batch size 1. Inference cost is measured for generating 25 samples. Fixed memory refers to the memory occupied by model parameters and optimizer states, and per-batch memory refers to the additional memory incurred by each training sample. All measurements are conducted on a single NVIDIA A100 GPU.}
\label{tab2}
\begin{tabular}{llcc}
\toprule
& & Proposed method & 3D Benchmark \\
\midrule
Model & Total parameters (M) & 526.35 & 1473.20 \\
\midrule
\multirow{3}{*}{Training} & Time: $\text{batch}=1$ (s) & 27.26 & 99.34\\
& Memory: fixed (GB) & 8.05 & 27.66 \\
& Memory: per batch (GB) & 1.47 & 27.60 \\
\midrule
\multirow{2}{*}{Inference} & Time: 25 samples (s) & 25.43 & 749.45 \\
& Memory: 25 samples (GB) & 6.68 & 75.10 \\
\bottomrule
\end{tabular}
\end{table}

\subsection{Computational efficiency}

A central motivation for the depth-as-channel formulation is to avoid the huge computational cost of training and deploying a native 3D diffusion model. To quantify this advantage, we compare the parameter count, training cost, and inference cost of the proposed method and the 3D benchmark under equivalent network configurations, where both methods use the same number of feature channels at each stage to ensure a fair comparison. The results are summarized in Table~\ref{tab2}.

The proposed method has 526.35M parameters, compared to 1473.20M for the 3D benchmark, a reduction of approximately 2.8 times. This difference reflects the fundamental advantage of operating with 2D convolutions under the depth-as-channel formulation rather than full 3D convolutions. Regarding training cost, the comparison is conducted at a batch size of 1 because the 3D benchmark, under the same network configuration, cannot accommodate larger batches within practical GPU memory limits. Under this setting, the proposed method requires 27.26~s per 100 iterations on a single NVIDIA A100 GPU, compared to 99.34~s for the 3D benchmark, a speedup of approximately 3.6 times. It is worth noting that diffusion models generally benefit from larger batch sizes for stable and effective training. When the batch size is increased, the training time gap between the two methods widens further. In terms of memory, the fixed overhead accounts for the model parameters and optimizer states that are independent of batch size, while the per-batch memory reflects the additional cost incurred by each training sample, including activations and gradients. The fixed memory consumption of the proposed method is 8.05~GB, compared to 27.66~GB for the benchmark. More critically, the per-batch memory is only 1.47~GB for the proposed method versus 27.60~GB for the benchmark, a factor of nearly 19. This large per-batch memory gap is what restricts the 3D benchmark to a batch size of 5 in the numerical experiments, while the proposed method can sustain much larger batch sizes without approaching memory limits.

The efficiency advantage becomes even more pronounced at inference time. Generating 25 samples takes 25.43~s with the proposed method, compared to 749.45~s for the 3D benchmark, a speedup of nearly 30 times. The corresponding memory consumption is 6.68~GB for the proposed method versus 75.10~GB for the benchmark, a reduction of more than 11 times. These results demonstrate that the depth-as-channel formulation not only maintains good VMB accuracy but also delivers substantial reductions in both training and inference cost, making the proposed method considerably more practical for large-scale 3D VMB.

\subsection{Limitations and future work}

The current implementation has one notable limitation regarding the structural constraint. In all experiments presented in this paper, the structural attribute is derived by convolving the true vertical reflectivity with a Ricker wavelet, which produces a high-quality image that faithfully captures the subsurface geometry. While this serves as an effective proof of concept for the proposed framework, such ideal structural information is not readily available in practice. Real exploration workflows require the structural constraint to be derived from recorded seismic data, where the image quality is degraded by acquisition artifacts, limited bandwidth, and incomplete illumination. A natural direction for future work is therefore to replace the convolution image with more realistic structural constraints. In our previous 2D work \citep{chengdiffvmbpart2}, we demonstrated that migration images derived from a smooth migration velocity model can serve as effective structural inputs, and extending this approach to the 3D setting is a straightforward next step toward field applicability.

A further direction concerns the direct use of 3D post-stack seismic data as the structural constraint for VMB. Post-stack data are routinely available in exploration settings and carry rich structural information in 3D. Incorporating 3D post-stack volumes as conditioning inputs would allow the proposed framework to operate more directly on acquired data, bypassing the need for a separate imaging step and potentially enabling a more tightly coupled workflow between seismic imaging and VMB.

\section{\textbf{Conclusions}}
We proposed a depth-progressive framework for 3D velocity model building (VMB) using 2D generative diffusion models. The core contribution is the depth-as-channel formulation, which maps the depth dimension of each 3D velocity patch to the channel dimension of a 2D U-Net, enabling volumetric VMB without the memory and computational burden of native 3D networks. The framework incorporates a recursive inference strategy, a classifier-free guidance scheme for optional well-log and structural constraints, and a depth-attenuated Gaussian blending scheme that accounts for the progressive accumulation of uncertainty toward greater depths. Numerical experiments on both in-distribution and out-of-distribution benchmark models demonstrated that the proposed method effectively propagates shallow velocity priors downward to construct 3D velocity models in a top-down manner, producing high-quality velocity volumes across a range of geological settings and conditioning scenarios, while achieving substantial reductions in parameter count, training time, and inference cost compared to a 3D benchmark of equivalent architecture. Currently, convolution images are used as a proxy for structural constraints. Future work will focus on incorporating more realistic structural inputs, such as migration images or 3D post-stack seismic volumes, to bring the framework closer to field applicability.

\section{Acknowledgments}
This publication is based on work supported by the King Abdullah University of Science and Technology (KAUST). The authors thank the DeepWave sponsors for their support. This work utilized the resources of the Supercomputing Laboratory at King Abdullah University of Science and Technology (KAUST) in Thuwal, Saudi Arabia.
\section{Code and Data Availability}

The source code, datasets, and pre-trained model weights for this work are publicly released to enable full reproducibility of the reported results.

\subsection{Code availability}

The complete implementation is hosted on GitHub at \url{https://github.com/DeepWave-KAUST/DiffVMB3D-pub}. The repository includes the core Python library in the \texttt{diffvmb3d/} directory, which implements the depth-as-channel 2D U-Net architecture, the Gaussian diffusion process with multi-source conditioning, and the depth-attenuated Gaussian blending for patch assembly. The training entry point \texttt{train.py} and the inference script \texttt{sample.py} are provided at the top level. The codebase builds on the improved denoising diffusion probabilistic models \citep{nichol2021improved} and extends it with classifier-free guidance for optional well-log and structural attribute conditioning. Environment setup instructions and usage examples are documented in the repository README.

\subsection{Dataset availability}

The training and testing datasets are archived on Zenodo at \href{https://doi.org/10.5281/zenodo.21206427}{DOI: 10.5281/zenodo.21206427} \citep{zenodo2026diffvmb3d}. The archive \texttt{train.zip} contains 5500 synthetic 3D velocity models stored as NumPy \texttt{.npz} files, each of size $128 \times 128 \times 128$ with three fields: the P-wave velocity model (\texttt{v3d}), the reflectivity model (\texttt{ref}), and the synthetic seismic image (\texttt{seis}) obtained by convolving the vertical reflectivity with a 60\,Hz Ricker wavelet at $\Delta t = 10^{-2}$\,s. The archive \texttt{test.zip} provides the four benchmark models evaluated in this manuscript (Overthrust, SEAM Arid, SEG/EAGE Salt, and Marmousi) as MATLAB \texttt{.mat} files with the same three fields. After downloading, users should extract the archives into the \texttt{dataset/} directory following the instructions in the repository README.

\subsection{Pre-trained model}

Pre-trained model weights are included in the same Zenodo record as the archive \texttt{trained\_model.zip}. The checkpoint corresponds to the EMA-averaged weights (decay rate 0.999) after 200\,000 training iterations on a single NVIDIA A100 GPU. After extracting the weights into the \texttt{checkpoints/} directory, the results reported in this manuscript can be reproduced by running:
\begin{verbatim}
python sample.py --use_ddim True --timestep_respacing ddim10 \
    --batch_size 50 --use_well True --use_ref True
\end{verbatim}
Different conditioning scenarios (unconditional, well-only, image-only, and well-plus-image) can be selected by toggling the \texttt{--use\_well} and \texttt{--use\_ref} flags. When well-log gradient guidance is enabled, users may additionally specify:
\begin{verbatim}
python sample.py --use_ddim True --timestep_respacing ddim10 \
    --batch_size 50 --use_well True --use_ref True \
    --use_wellguide True --scale_factor 20
\end{verbatim}

\bibliographystyle{unsrtnat}
\bibliography{references}

\end{document}